\documentclass[lettersize,journal]{IEEEtran}
\usepackage{array}
\usepackage{textcomp}
\usepackage{url}
\usepackage{verbatim}
\usepackage{graphicx}

\usepackage{cite}
\usepackage{tikz}
\usepackage[breaklinks=true]{hyperref}
\hypersetup{
     colorlinks=true,
     linkcolor=black,
     filecolor=black,
     citecolor = black,      
     urlcolor=black,
}
\usepackage{diagbox}
\usepackage{booktabs}
\usepackage{amsmath}
\usepackage{epsfig,graphicx,subcaption,tabularx,relsize,steinmetz}
\usepackage[nameinlink]{cleveref}
\usepackage{aliascnt}
\crefname{equation}{}{}
\crefname{algorithm}{Algorithm}{Algorithms}
\makeatletter

\def\endineq{\eqno \hbox{\@eqnnum}$$\@ignoretrue}
\makeatother
\crefname{figure}{Fig.}{Figs.}
\crefname{table}{Table}{Tables}
\usepackage{algorithm,algpseudocode,float}


\makeatletter
\newcommand{\multiline}[1]{%
  \begin{tabularx}{\dimexpr\linewidth-\ALG@thistlm}[t]{@{}X@{}}
    #1
  \end{tabularx}
}
\makeatother

\usepackage{balance}
\begin{document}
\title{Optimal Trajectory Planning for Connected and Automated Vehicles in Lane-free Traffic with Vehicle Nudging}
\author{Venkata Karteek Yanumula, Panagiotis Typaldos, Dimitrios Troullinos, Milad Malekzadeh, Ioannis Papamichail, Markos Papageorgiou,~\IEEEmembership{Life~Fellow,~IEEE}
\thanks{Manuscript created May, 2022.}
\thanks{This research received funding from the European Research Council under the EU Horizon 2020 Programme / ERC Grant no. 833915, project TrafficFluid, see: \href{https://www.trafficfluid.tuc.gr}{\tt\small \nolinkurl{https://www.trafficfluid.tuc.gr}}}
\thanks{All the authors are with Dynamic Systems and Simulation Laboratory, Technical University of Crete, Chania, Crete, 73100, Greece, e-mail: {\tt\small \{karteek, ptypaldos, dtroullinos, mmalek, ipapa, markos\}@dssl.tuc.gr}}% <-this % stops a space
\thanks{Venkata Karteek Yanumula is with Electrical and Instrumentation Engineering Department, Thapar Institute of Engineering and Technology, Patiala, India.}
\thanks{Markos Papageorgiou is also with Faculty of Maritime and Transportation, Ningbo University, Ningbo, China.}
}

\markboth{Published in IEEE Transactions on Intelligent Vehicles. This is a full version of the same.}
{}

\maketitle

\begin{abstract}
The paper presents a movement strategy for Connected and Automated Vehicles (CAVs) in a lane-free traffic environment with vehicle nudging by use of an optimal control approach. State-dependent constraints on control inputs are considered to ensure that the vehicle moves within the road boundaries and to prevent collisions. An objective function, comprising various weighted sub-objectives, is designed, whose minimization leads to vehicle advancement at the desired speed, when possible, while avoiding obstacles. A nonlinear optimal control problem (OCP) is formulated for the minimization of the objective function subject to constraints for each vehicle. A computationally efficient Feasible Direction Algorithm (FDA) is called, on event-triggered basis, to compute in real-time the numerical solution for finite time-horizons within a Model Predictive Control (MPC) framework. The approach is applied to each vehicle on the road, while running simulations on a lane-free ring-road, for a wide range of vehicle densities and different types of vehicles. From the simulations, which create myriads of driving episodes for each involved vehicle, it is observed that the proposed approach is highly efficient in delivering safe, comfortable and efficient vehicle trajectories, as well as high traffic flow outcomes. The approach is under investigation for further use in various lane-free road infrastructures for CAV traffic.
\end{abstract}


\section{Introduction}
CAVs (Connected Automated Vehicles) are expected to revolutionize the transportation sector, improving on traffic safety and efficiency by overcoming the limitations of human driving. The current driving pattern followed by human drivers is resulting in underuse of the road infrastructure, excessive fuel consumption and reduced safety \cite{singh2015}. Human perception of the surroundings is limited compared to the wide range of sensors \cite{enrique2019} that are available onboard CAVs. In these circumstances, accidents and congestion are regular plaguing occurrences across the world, calling for comprehensive and radical solutions. One potential solution would be introducing CAVs on the roads with appropriate movement strategies. However, due to the strongly dynamic nature of road traffic in both longitudinal and lateral directions, complex algorithms, like optimal model predictive control, nonlinear feedback control, reinforcement learning, and artificial intelligence, are necessary \cite{david2016,karteek2021,iasson2021,dimitri2021,Fei2022-1}. With appropriate sharing of information via Vehicle-to-Vehicle (V2V) and Vehicle-to-Infrastructure (V2I) communication, CAVs may be enabled to drastically alleviate the problems of safety and congestion \cite{seng2019,diakaki2015}.  

Traffic lanes were introduced to aid the human driving, as they minimize the need for lateral and rear monitoring; albeit with an exception during lane-changing, which calls for simultaneous multi-direction monitoring, prediction of multiple vehicle movements and fast decision making. Around 10\% of all road accidents occur while performing the lane-change by human drivers \cite{rahman2013}. Besides, due to diverse vehicle dimensions, lanes are designed to fit the widest vehicles in regular traffic, which are big trucks. Thus, a great deal of lateral space is left unused when smaller vehicles are using the same lane structure. At traffic level, lane-changes lead to capacity losses and are known to trigger traffic breakdown at critical traffic conditions. 

Automated driving systems that mimic human driving have been using several approaches to generate the lane-changing maneuvers \cite{david2016-2,bai2018}, which are complex procedure even for CAVs. For example, optimal control techniques within a model predictive control (MPC) scheme have been proposed for lane-based trajectory planning, including lane-changing \cite{Makantasis2018,panos2022}. However, with the large amount, quality and range of perception data available from various sensors, with V2V and V2I communications, and with the development of efficient trajectory planning algorithms, CAVs may drive without consideration of lane structure on the roads. Moreover, several other traffic rules, designed for human drivers, can be relaxed for CAVs, without compromising on safety, while improving the efficiency of traffic.

For future traffic, dominated by fully automated vehicles, there is no apparent reason to necessarily employ human-inspired driving restrictions, such as lane-based driving or accounting of downstream vehicles only (as in conventional car-following). Some works (e.g. \cite{amouzadi2022}) proposed CAV lane-free crossing of urban junctions, see \cite{sekeran2022} for a brief historical review. A general traffic paradigm is proposed in \cite{markos2021}, characterized by lane-free CAV driving and vehicle ``nudging". In fact, CAVs have sufficient sensors to continuously scan the $360^\circ$ surroundings reliably and fast. Thus, the CAVs, backed with appropriate trajectory planning algorithms, may move on a lane-free road, like particles constituting an artificial fluid, whereby vehicle movement may be influenced not only by other vehicles in front of them (as with human driving), but also by vehicles around and behind, something that is referred to as nudging and seems to have beneficial implications for traffic flow capacity and stability \cite{iasson2020}. Lane-free structure improves the usability of lateral road space by the vehicles and has additional potential benefits, such as increased capacity with incremental road widening over limited road length, as well as flexible (in space and time) incremental sharing of the total road capacity among the two traffic directions in real-time \cite{milad2021}. Using appropriate algorithms, the traffic on motorways, highways, arterials, and urban roads could be freed from the obligation to drive on lanes and regain the lost capacity without compromising on safety. The discontinuous lateral displacements from one lane to another become obsolete, and smooth lateral vehicle movement may be employed for overtaking and other driving tasks.

This article develops a nonlinear constrained Optimal Control Problem (OCP), with both fixed and state-depended bounds on control inputs, for the movement of CAVs in lane-free traffic involving vehicle nudging. Thanks to the lane-free vehicle movement, real-valued displacements on the lateral direction are considered in the OCP, in contrast to the introduction of discrete variables engaged in most lane-based optimal trajectory planning approaches. 

Due to the highly dynamic nature of traffic and the related uncertainty increase with time, relatively short planning time-horizons (8 s) are considered within an MPC framework with receding time-horizon. MPC is a time-tested approach with applications in various fields \cite{mayne2014}. In the present study, an efficient Feasible Direction Algorithm (FDA) \cite{fda} is employed to solve the OCP problem within an event-based MPC framework \cite{yasaman2018}. It should be noted that the optimal trajectories of 8-s duration are computed in 12.4 ms of CPU time on average, which allows for convenient application in real-time.

A lot of developments are ongoing in research and industry to enable reliable V2I and V2V information sharing in real-time \cite{jin2019,Fei2022-2,Fei2022-3}. In the present work, it is assumed that there is sufficient communication among adjacent vehicles within a limited range, both upstream and downstream of each ego vehicle (EV), which enables vehicles to share information related to their position, speed, and the latest calculated optimal trajectories. 

In the present work, the OCP is solved independently for hundreds of vehicles, driving on a ring-road, based on event-triggered MPC for each vehicle, in emulated real-time. This allows for investigation of both vehicle-level and traffic-level implications of the proposed vehicle movement strategy. It is found that a fundamental diagram (flow-density curve), with a typical inverse-U shape, emerges at traffic level. It is also observed that the vehicles move efficiently, smoothly and safely over the full range of possible vehicle densities, while maintaining passenger convenience. Emerging flow values are significantly higher than in conventional lane-based traffic.

A preliminary version of the approach with limited evaluation results was reported in \cite{karteek2021}. The concept of lane-free roads and vehicle trajectory planning using optimal control within a nonlinear MPC framework is also considered in \cite{levy2021}, which features similarities, but also significant differences, compared to \cite{karteek2021} and the present work, with respect to the OCP formulation, the numerical solution, the considered MPC procedures and the application focus. 

The major contributions of the article are as listed below;
\begin{itemize}
\item Development and efficient numerical solution of an optimal control approach for vehicle movement on a lane-free roads with nudging.
\item Comprehensive testing of the trajectory planning approach for hundreds of vehicles, involved in myriads of driving circumstances.
\item Traffic-level simulation-based evaluation for a wide range of densities, demonstrating sensible improvements.
\end{itemize}
The rest of the article is organized as follows: the kinematic model of the vehicle dynamics for each EV is discussed in \Cref{veh_model}, while constraints on lateral and longitudinal accelerations are designed in \Cref{sdbc}. The OCP formulation, numerical solution and employed MPC features are detailed in \Cref{ocp}. Simulation environment, vehicle-level and emerging traffic-level results and related discussion are given in \Cref{sim}. Finally, the concluding remarks are provided in \Cref{conc}.

\section{Vehicle Model and Constraints }
This section describes the mathematical vehicle model, along with the fixed and state-dependent control constraints.
\subsection{Vehicle Model}\label{veh_model}
A discrete-time vehicle model is considered, governed by discrete-time control inputs. In the present work, vehicles are moving on an unfolded straight ring-road, with small lateral speed compared to longitudinal speed. Therefore, the heading angles of the vehicles are small, and, consequently, each vehicle is represented by two decoupled double integrators in discrete-time, one for each of the longitudinal and lateral directions. As a precaution, a specific term is considered in the objective function to discourage infeasible heading angles (see \Cref{long_cpl}). Furthermore, the lateral movement uses real-valued states thanks to the lane-free road structure, which allows for the vehicles to be positioned within the road boundaries continuously in the lateral direction. The mathematical model of each EV in the form of state equations is:

\begin{subequations}\label{eq1}
\begin{align}
x_1(k+1) &= x_1(k) + Tx_3(k) + \frac{1}{2} T^2 u_1(k) \label{eq1a}\\
x_2(k+1) &= x_2(k) + Tx_4(k) + \frac{1}{2} T^2 u_2(k) \label{eq1b}\\
x_3(k+1) &= x_3(k) + T u_1(k) \label{eq1c}\\
x_4(k+1) &= x_4(k) + T u_2(k) \label{eq1d}
\end{align}
\end{subequations}
where $x_1$ and $x_2$ are longitudinal and lateral positions, respectively; $x_3$ and $x_4$ are speeds in longitudinal and lateral directions, respectively; $k$ is the discrete-time integer index; and $T$ is the discrete-time step. The index $k$ relates the step $T$ and continuous time $t$ through $t=kT$. The control inputs $u_1$ and $u_2$ are the vehicle accelerations in longitudinal and lateral directions, respectively. Both control inputs are bounded, as discussed in \cref{sdbc}. It should be noted that the state equation above is the exact discrete-time version of the corresponding continuous-time double-integrator movement equations, if the control inputs are held constant for the duration $T$ of each time interval.

\subsection{Fixed and State-dependent Bounds on Control}\label{sdbc}
In view of limited vehicle capabilities, passenger comfort considerations, and the need to respect the road boundaries, the control inputs (longitudinal and lateral accelerations) must be bounded appropriately.

\subsubsection{Longitudinal Bounds on Control}\label{sdbc-1}
A state-dependent lower limit on the longitudinal acceleration is obtained from the requirement for non-negative speed at the next time-step $k+1$; i.e., we have the requirement

\begin{equation}\label{sdbc-eq1}
x_3(k+1) \geq 0.
\end{equation}
Using \cref{eq1c}, \cref{sdbc-eq1} yields $u_1(k) \geq -\frac{1}{T}x_3(k)$, hence a state-dependent lower limit on longitudinal acceleration is $U_\text{min1}(k)= -\frac{1}{T}x_3(k)$. In addition, the vehicle has limited deceleration capabilities, hence a constant limit $U_\text{min1}$ is also considered, and the overall lower limit of the longitudinal acceleration is 

\begin{equation}\label{sdbc-eq2}
u_\text{min1}(k) = \text{max}\left\{U_\text{min1}(k),U_\text{min1}\right\}.
\end{equation}
Regarding the upper limit on the longitudinal acceleration, there are also two kinds of bounds to account for, a constant bound and a state-dependent bound. The constant upper bound $U_\text{max1}$  reflects the vehicle's acceleration capabilities; this bound is considered constant, but it could be readily expressed as a function of longitudinal speed for more realism. 

On the other hand, the state-dependent upper bound is used only in rare emergency situations, as described later (\Cref{mpc}). For now, it suffices to mention that, in such cases, an upper bound on the longitudinal acceleration is obtained from the requirement of following the obstacle vehicle downstream at a prespecified distance $\epsilon$. A corresponding longitudinal position limit $\widehat{x}_1(k)$ behind the rear bumper of the obstacle is defined, and the EV must not cross this limit. This longitudinal position limit may be viewed as a moving boundary; Indeed, the longitudinal position limit moves with the vehicle dynamics of the obstacle. Consider the following dynamics for the moving boundary in the form of state equations, where all $\widehat{\cdot}$ variables have similar meanings as in \cref{eq1},

\begin{subequations}\label{sdbc-eq3}
\begin{align}
\widehat{x}_1(k+1) &= \widehat{x}_1(k) + T\widehat{x}_3(k) + \frac{1}{2} T^2 \widehat{u}_1(k) \label{sdbc-eq3a}\\
\widehat{x}_3(k+1) &= \widehat{x}_3(k) + T \widehat{u}_1(k). \label{sdbc-eq3b}
\end{align}
\end{subequations}
The EV, positioned behind the moving boundary at $x_1(k)$ and $x_1(k+1)$ should not cross the moving boundary at time step $k+2$ i.e. we must have
\begin{align}\label{sdbc-eq4}
x_1(k+2) \leq \widehat{x}_1(k+2).
\end{align}
Note that, if the EV reaches the moving boundary, and constraint \cref{sdbc-eq4} is activated, then the longitudinal speed of the EV should be equal to that of the moving boundary, otherwise violation of \cref{sdbc-eq4} cannot be avoided eventually. By substituting \cref{eq1a,eq1c,sdbc-eq3} in \cref{sdbc-eq4} and with some rearrangement, the following upper limit on the longitudinal acceleration is obtained
\begin{align}\label{sdbc-eq5}
u_1(k) \leq -\frac{1}{T^2}\left[x_1(k) - \widehat{x}_1(k)\right] - \frac{3}{2T}\left[x_3(k)-\widehat{x}_3(k)\right] + \widehat{u}_1(k).
\end{align}
Defining the tracking errors 
\begin{subequations}\label{sdbc-eq6}
\begin{align}
e_1(k) &= x_1(k) - \widehat{x}_1(k)\label{sdbc-eq6a}\\
e_2(k) &= x_3(k) - \widehat{x}_3(k) \label{sdbc-eq6b}
\end{align}
\end{subequations}
the right-hand side of \cref{sdbc-eq5} may be viewed as a tracking controller. In fact, if the upper limit in \cref{sdbc-eq5} is activated and continuously applied, the EV's position and speed reach the moving boundary's position and speed in exactly two time-steps, i.e., the tracking controller exhibits a dead-beat behaviour. However, the magnitude of the longitudinal acceleration required for such a dead-beat behaviour could be very high and might violate the constant acceleration bounds. 

The required magnitude of the longitudinal acceleration can be moderated by choosing appropriate feedback gain values $0<K_\text{long1}\leq 1/T^2$ and $0<K_\text{long2} \leq 3/(2T)$. In place of \cref{sdbc-eq5}, consider the maximum longitudinal control input of EV as

\begin{align} \label{sdbc-eq7}
u_\text{max1}(k)=-K_\text{long1} e_1(k) - K_\text{long2} e_2(k) + \widehat{u}_1(k).
\end{align}
The choice of $K_\text{long1}$ and $K_\text{long2}$ should result in aperiodic tracking of the set-points, when the limit is active, such that the EV reaches the required position and speed asymptotically, reducing the tracking error to zero without any overshooting. To choose appropriate gain values for this behaviour, the following error dynamics are obtained using \cref{eq1a,eq1c,sdbc-eq3,sdbc-eq7}, when the limit is active, i.e., when $u_1(k) = U_\text{max1}(k)$,

\begin{align}\label{sdbc-eq8}
\begin{bmatrix}
e_1(k+1) \\ e_2(k+1)
\end{bmatrix} &= \begin{bmatrix}
1 - \frac{1}{2}K_\text{long1}T^2 & T - \frac{1}{2}K_\text{long2}T^2 \\
-K_\text{long1}T & 1-K_\text{long2}T
\end{bmatrix}\begin{bmatrix}
e_1(k) \\
e_2(k)
\end{bmatrix}.
\end{align}
For stability and aperiodic behaviour, the gain values $K_\text{long1}$ and $K_\text{long2}$ should ensure that the roots of the characteristic equation of \cref{sdbc-eq8} are inside the unit circle (for stability) and non-negative real (for aperiodic behaviour). The characteristic equation for \cref{sdbc-eq8} is given by

\begin{align}\label{sdbc-eq9}
z^2+ & \left(\frac{K_\text{long1} T^2}{2}+TK_\text{long2} - 2 \right) + \nonumber \\ &\left(\frac{K_\text{long1}T^2}{2} -TK_\text{long2}+1 \right) = 0.
\end{align}
To keep the design procedure simple, we choose $K_\text{long2}=2\sqrt{K_\text{long1}}-K_\text{long1}T/2$, for which the roots of the characteristic equation are non-negative real and identical. The resulting feedback values (longitudinal accelerations) are kept moderate by appropriately choosing the gain value $K_\text{long1}$  within the mentioned range. 

The emergency planning to avoid longitudinal collision will always result in an upper limit on acceleration that is less than that of constant bound $U_\text{max1}$, which is used in regular planning. Thus, the upper limit on the longitudinal acceleration is either $u_\text{max1}(k) = U_\text{max1}$ or $u_\text{max1}(k) = U_\text{max1}(k)$ for regular or emergency planning, respectively.

\subsubsection{Lateral Bounds on Control}\label{sdbc-2}
The main requirement on the lateral direction is for the vehicle to stay inside the road boundaries, or to drive exactly on a road boundary, without violating them. In addition, some imaginary boundaries may be imposed in the rare emergency cases. In both cases, the boundaries are stationary straight lines in this study, while other shapes of road boundaries are currently under investigation. The imaginary boundaries in the emergency cases are defined so as to prevent collisions on lateral direction, and related details are provided in \Cref{mpc}. 

Using a similar approach as in \Cref{sdbc-1}, the EV, positioned within the road boundaries at time steps $k$ and $k+1$, should also stay within the road boundaries at time step $k+2$, i.e., we must have

\begin{align}\label{sdbc-eq10}
\widetilde{x}_2(k+2) \leq x_2(k+2) \leq \widehat{x}_2(k+2)
\end{align}
where $\widehat{x}_2$ is the lateral position of the left road boundary (equal to the road width minus half of the EV width); and $\widetilde{x}_2$ is the lateral position of the right road boundary (equal to half of the EV width). Note that, if the vehicle actually reaches the left or right road boundary, i.e., if the left or right constraint above is activated, then we must have for the lateral vehicle speed $x_4\left(k+2\right)=0$, as otherwise the vehicle would eventually exit the road. Thus, compared with \Cref{sdbc-1}, we have, for fixed straight boundaries, the simpler case that$\widehat{u}_2\left(k\right)=0$, $\widehat{x}_4\left(k+1\right)=\widehat{x}_4\left(k\right)=0$, $\widehat{x}_2\left(k+1\right)=\widehat{x}_2\left(k\right)=\widehat{x}_2$, $\widetilde{u}_2\left(k\right)=0$, $\widetilde{x}_4\left(k+1\right)=\widetilde{x}_4\left(k\right)=0$, and $\widetilde{x}_2\left(k+1\right)=\widetilde{x}_2\left(k\right)=\widetilde{x}_2$. By replacing \cref{eq1b,eq1d} in \cref{sdbc-eq10} and after some rearrangements, we have that the above constraints are equivalent to the following state-dependent inequalities for the lateral acceleration 

\begin{align}\label{sdbc-eq11}
-\frac{1}{T^2}[x_2(k)&-\widetilde{x}_2(k)]-\frac{3}{2T}x_4(k) \leq u_2(k) \nonumber \\ &\leq -\frac{1}{T^2}\left[x_2(k)-\widehat{x}_2(k)\right]-\frac{3}{2T}x_4(k)
\end{align}
Similarly to \cref{sdbc-eq5}, either side of \cref{sdbc-eq11} can be considered, if taken as an equality, as a respective dead-beat controller that drives the EV towards the corresponding road boundary and its lateral speed to zero in two time steps. Also here, the magnitudes of the resulting lateral accelerations produced by such a controller may take unrealistically high values and may cause discomfort to the passengers. Again, we can design more moderate controller gains, such that the lateral acceleration is feasible and comfortable. Thus, similarly to the generalization in \cref{sdbc-eq7}, we have the state-dependent bounds on lateral acceleration given by

\begin{subequations}\label{sdbc-eq12}
\begin{align}
U_\text{min2}(x_2(k),x_4(k)) = -K_\text{lat1}[x_2(k) - \widetilde{x}_2(k)] - K_\text{lat2}x_4(k) \label{sdbc-eq12a} \\
U_\text{max2}(x_2(k),x_4(k)) = -K_\text{lat1}[x_2(k) - \widehat{x}_2(k)]- K_\text{lat2}x_4(k) \label{sdbc-eq12b}
\end{align}
\end{subequations}
where $0 < K_\text{lat1} \leq 1/T^2$ and $0 < K_\text{lat2} \leq 3/(2T)$ are feedback controller gains, chosen by following a similar procedure as in \Cref{sdbc-1} and setting $K_\text{lat2}=2\sqrt{ K_\text{lat1}}- K_\text{lat1}T/2$ for two identical non-negative real poles, so that we achieve asymptotic behaviour with moderate lateral accelerations. Clearly, \cref{sdbc-eq12} apply to either the road boundaries or to imaginary lateral boundaries (in emergency cases), with $\widehat{x}_2$ and $ \widetilde{x}_2\left(k\right)$ set accordingly.

It is interesting to note that, in a lane-free traffic environment, guaranteeing that a vehicle will not violate the road boundaries is more important than in a lane-based environment, for two reasons. First, lane-based vehicles are typically guided around the middle of each lane, and therefore the risk of departing from the road is less serious. Second, in lane-free driving, there is a strong interest, e.g., at dense traffic, to let some vehicles drive \emph{exactly} on a (left or right) road boundary (of course without ever violating it), so as to maximize the usage of the available road width. Note that exact road boundary tracking, under lateral pressure (nudging) by laterally adjacent vehicles, e.g. due to high traffic density, is hardly possible by use of related penalty or barrier terms in the optimisation objective, as in lane-based driving \cite{panos2022}. 

These two endeavours of lane-free driving are fulfilled via the above two boundary controllers, one for the left and another for the right road boundary, which deliver appropriate upper and lower bounds, respectively, for the lateral acceleration. As long as such a bound is not activated, the vehicle drives according to other optimal control criteria (see \Cref{obj}); if one of the bounds is activated, it was designed above to navigate the vehicle asymptotically to the corresponding road boundary and then have it driving exactly on the road boundary, for as long as the lateral acceleration is activating that bound. 

For consistency of notation, we define the lower and upper limits as $u_\text{min2}\left(k\right)=U_\text{min2}(k)$ and $u_\text{max2}\left(k\right)=U_\text{max2}(k)$ respectively. 

In summary, the bounds on accelerations given in \cref{sdbc-eq2,sdbc-eq7,sdbc-eq12} are expressed by the following state-dependent constraints, to be considered in the OCP

\begin{subequations}\label{sdbc-eq13}
\begin{align}
h_1&=[u_1(k) - u_\text{max1}(k)][u_1(k) - u_\text{min1}(k)] \leq 0 \label{sdbc-eq13a} \\
h_2&=[u_2(k) - u_\text{max2}(k)][u_2(k) - u_\text{min2}(k)] \leq 0. \label{sdbc-eq13b}
\end{align}
\end{subequations}

\section{Optimal Control Problem (OCP)}\label{ocp}
The OCP considered in this work minimizes the weighted sum of multiple sub-objectives for the EV advancement in an efficient, safe and comfortable manner. This section details the objective function, followed by the formulation of the discrete-time OCP to be solved numerically. The section also presents the FDA for numerical solution of the OCP and the MPC procedure employed in real time.

\subsection{Objective Function}\label{obj}
The objective function is composed by several sub-objectives that are designed to mitigate discomfort for the passengers while applying accelerations on longitudinal and lateral directions, to reach desired speeds, to avoid dynamic obstacles and more. For application of the numerical algorithm, each of these sub-objectives must be continuous and differentiable.

\subsubsection{Fuel Consumption and Passenger Comfort}\label{fuel_cons}
Mitigation of fuel consumption results in accordingly reduced emissions and environment-friendly CAV traffic. This is achieved by a quadratic cost term of longitudinal acceleration $\left(u_1\left(k\right)\right)^2$ as demonstrated in \cite{panos2020}. In addition, quadratic cost terms of longitudinal and lateral accelerations $\left(u_1\left(k\right)\right)^2$ and $\left(u_2\left(k\right)\right)^2$, respectively, result in moderate and smooth accelerations and thereby improved passenger comfort.

\subsubsection{Desired Speed}\label{des_sp}
Reaching desired speeds on both longitudinal and lateral directions is essential for vehicle advancement and reaching of an exit. The road considered in this study is an unfolded ring-road, requiring a positive longitudinal desired speed and a zero lateral desired speed. The quadratic cost term $\displaystyle \left[x_3(k) - v_{d_1}\right]^2$ with longitudinal desired speed $v_{d_1}$ results in the EV trying to reach its desired speed. Clearly, vehicles moving near their respective desired speeds reduce the travel time of passengers. Unnecessary lateral movements can reduce traffic efficiency, hence the quadratic cost term $\displaystyle \left[x_4(k)-v_{d_2}\right]^2$, here with $v_{d_2}=0$ as lateral desired speed, mitigates lateral movement. In ongoing work with highways including on-ramps and off-ramps, non-zero lateral desired speeds may be appropriately used for efficient vehicle merging on or departing from the mainstream.

\subsubsection{Obstacle Avoidance}\label{obst_avoid}

For obvious safety reasons, the EV should avoid collisions with other vehicles that are treated as moving obstacles. This is an important requirement that must be reflected in a carefully designed sub-objective with appropriate consideration of the novel aspects stemming from the lane-free vehicle movement, as well as from the possibility of vehicle nudging. 

To avoid collisions, while generating the trajectory of the EV, it is necessary that the current states (position and speed) of surrounding vehicles are available via the EV sensors or via V2V communication. In addition, each vehicle shares its generated trajectory to surrounding vehicles. Naturally, short-term prediction of the movement of surrounding vehicles may have limited accuracy, and this may trigger a new path generation, see \Cref{mpc} for details.

While solving the OCP for finite-time horizons, an \textit{interaction zone} (IZ) is considered, and all the other vehicles inside the IZ are treated as obstacles. The said IZ for each planning horizon has two equal-length parts, one downstream and the other upstream of the EV, the latter serving the purpose of enabling vehicle nudging, i.e., the possibility for upstream vehicles to influence the movement of vehicles downstream. The length of each of these parts is equal to the product of longitudinal desired speed times the planning horizon duration. Since vehicles are assumed to broadcast their trajectory decisions to other surrounding vehicles, each EV has such trajectory information of all the obstacles in its IZ.

Maintaining longitudinal and lateral safety gaps from surrounding obstacles is necessary in lane-free traffic. In lane-based driving, a longitudinal safety gap to the front vehicle on the same lane, which is proportional to the EV speed, is generally employed \cite{michail2020}, and this is termed as constant time-gap policy. In this work, an imaginary positive-valued, smooth, longitudinally and laterally symmetric ellipsoid hemisphere around each obstacle is carefully designed, which determines the cost incurred if the EV is positioned within the ellipsoid range. This cost is aimed to discourage the EV from intruding the ellipsoid range, i.e., from approaching too closely (or colliding with) the obstacle. While designing the ellipsoid hemisphere, time-gap like safety parameters $\left(\omega_1,\omega_2\right)$ are considered in longitudinal and lateral directions, respectively, along with the obstacle's corresponding physical dimensions. Assuming $n$ obstacles inside the IZ, we define for the $i^{\mathrm{th}}$  obstacle the longitudinal and lateral positions $\left(o_{i1},o_{i2}\right)$ and speeds $\left(o_{i3},o_{i4}\right)$. 

The length and position (centre) of the ellipsoid in the longitudinal direction are specified to satisfy the following requirements:

\begin{itemize}
\item Physical dimensions of both ego and obstacle vehicles should be completely covered by the ellipsoid, even for zero speed. We define $L_i=\mu_x\left(l_e+l_{oi}\right)$, where $l_e$ and $l_{oi}$ are the lengths of ego and obstacle vehicles, respectively, and $\mu_x$ is a coefficient. When both ego and obstacle vehicles are at zero speed, the length of the longitudinal axis of the ellipsoid equals $L_i$, hence $\mu_x$ ensures that vehicles are not bumper-to-bumper at stillstand.
\item If the EV is positioned behind the obstacle, a safety gap, proportional to its current speed, should be considered (in addition to $L_i$), according to the constant time-gap policy. Thus, a safety gap equal to $\omega_1x_3$ should be maintained in front of the EV.
\item If the EV is positioned ahead of the obstacle, a safety gap proportional to the current speed of the obstacle should be considered (in addition to $L_i$). Thus, a safety gap equal to $\omega_1o_{i3}$ should be maintained behind the EV. This is useful for preventing the EV from cutting-in dangerously close in front of an obstacle. It is interesting to note that the consideration of a safety gap not only in front, but also behind the EV is creating a longitudinal ``nudging" to the EV due to a moving obstacle behind it. More generally, due to this gap, vehicle driving is not influenced only by obstacles in front, as in manual driving or in most lane-based automated vehicle concepts, but also by obstacles behind them, something that gives rise to structurally new traffic flow characteristics at the macroscopic level \cite{iasson2020,iasson2022}.
\item Due to the different length of the safety gaps in front and behind the EV, which depend on the EV speed and the following obstacle speed, respectively, the positioning of the centre of the symmetric ellipsoid in the longitudinal direction depends on the speed difference of the ego and obstacle vehicles. The ellipsoid centre coincides with the obstacle centre if both vehicles have the same speed; but the ellipsoid centre is shifted longitudinally, to reflect properly the above safety gap specifications, if the speeds of the ego and obstacle vehicles are different.
\end{itemize}

All mentioned requirements are fulfilled, if the length of the longitudinal ellipsoid axis is equal to $d_1=L_i+\omega_1x_3+\omega_1o_{i3}$, while the ellipsoid's longitudinal centre is positioned at

\begin{equation}\label{ocp-eq1}
\delta_{oi} = o_{i1} - \omega_1\left(x_3-o_{i3}\right)/2
\end{equation} 
Consideration of the lateral direction in collision avoidance is of paramount importance in lane-free driving, in contrast to lane-based driving, where lateral vehicle collisions are not an issue (except in lane changing manoeuvres). In the lateral direction, the safety gap between the vehicles is necessary only when they are approaching each other. The safety gap considered is proportional to the absolute value of speed difference and is equal to $\omega_2\left|x_4-o_{i4}\right|$, when ego and obstacle vehicles are approaching each other, else it is equal to 0. A smooth approximation of the lateral safety gap, accounting for the mentioned specifications, is 

\begin{align}\label{ocp-eq2}
d_2 = W_i + &\omega_2\Big[\tanh(o_{i2}-x_2)\left(x_4-o_{i4}\right) \nonumber \\ &+ \sqrt{\left[\tanh(o_{i2}-x_2)\left(x_4-o_{i4}\right)\right]^2+\epsilon_w}\Big] 
\end{align}
where $\epsilon_w$ has a small value and $W_i=\mu_y\left(w_e+w_{oi}\right)$ with $w_e$ and $w_{oi}$ the widths of ego and obstacle vehicles, respectively. A coefficient $\mu_y$ slightly greater than 1 is considered to ensure that the corners of the (rectangular) vehicles are sufficiently covered by the ellipsoid, so as to suppress corner-corner collisions. There is an interest that $\mu_y$ is not much bigger than 1, as this would waste lateral space and hence reduce the resulting traffic flow capacity. The lateral midpoint of the ellipsoid coincides with the centre of the obstacle in lateral direction. Note that the lateral dimension of the ellipsoid, in particular the lateral safety gaps, create also a lateral ``nudging" to the EV due to other, longitudinally close, vehicles around it, if their lateral distance from the EV is diminishing, i.e., if both vehicles are approaching each other laterally.

The ellipsoid function, which covers sufficiently physical vehicle dimensions and safety gaps, is essentially a potential function, whose value smoothly increases as the EV approaches to the obstacle (or vice-versa), and acts as a collision avoidance term in the objective function, separate for each obstacle within the EV’s IZ. Besides the need for obstacle avoidance, the shape of the ellipsoid should satisfy some additional requirements that stem from our interest in achieving high traffic flow capacity, efficient vehicle manoeuvring, as well as from the gradient-based nature of the numerical solution algorithm. Specifically, we have the following two requirements for the ellipsoid function:

\begin{enumerate}
\item[(i)] 	The shape of the ellipsoid's iso-cost curves (contours) around the obstacle should be sufficiently rectangular to cover the obstacle's corners (particularly at low speeds where safety gaps are short), mitigating the need for larger coefficient $\mu_y$, which would waste lateral space and capacity. On the other hand, the contour shape should be sufficiently ``curvy" at both longitudinal edges, so that the EV, guided by the ellipsoid gradient, may ``slide" around an encountered slower obstacle in front of it; rather than being stuck behind the obstacle due to a quasi-straight contour edge.
\item[(ii)]	The ellipsoid's three-dimensional shape should be sufficiently steep, i.e., its gradient should be sufficiently far from zero, at all points of its range, to facilitate faster convergence of the numerical solution algorithm to collision-free trajectories.
\end{enumerate} 
While many functions may satisfy the above requirements, the function selected here for the ellipsoid reads

\begin{align}\label{ocp-eq3}
c_i(\boldsymbol{x},\ \boldsymbol{o}_i) = &\left\{1-\tanh\left[\left(\frac{x_1 - \delta_{oi}}{0.5d_1}\right)^{p_1}+\left(\frac{x_2 - o_{i2}}{0.5d_2}\right)^{p_2}\right]\right\} \nonumber \\ & + \left\{\left[\left(\frac{x_1 - \delta_{oi}}{0.25d_1}\right)^{p_3}+\left(\frac{x_2 - o_{i2}}{0.25d_2}\right)^{p_4}\right]^{p_5} + 1 \right\}^{-1}
\end{align}
where $\boldsymbol{x}$ and $\boldsymbol{o}_i$ are state vectors comprising the four state variables of the ego and obstacle-$i$ vehicles, respectively. The exponents $p_1$ to $p_5$ influence the ellipsoid shape. Specifically, $p_1$ to $p_4$ take positive even integer values, while $p_5$ is a positive integer. The function described in \cref{ocp-eq3} is a sum of two ellipsoids and it is an extended version of the ellipsoid employed in \cite{karteek2021}. 

\begin{figure}[!t]
\centering
\hspace{-8mm}
	\includegraphics[scale=0.23]{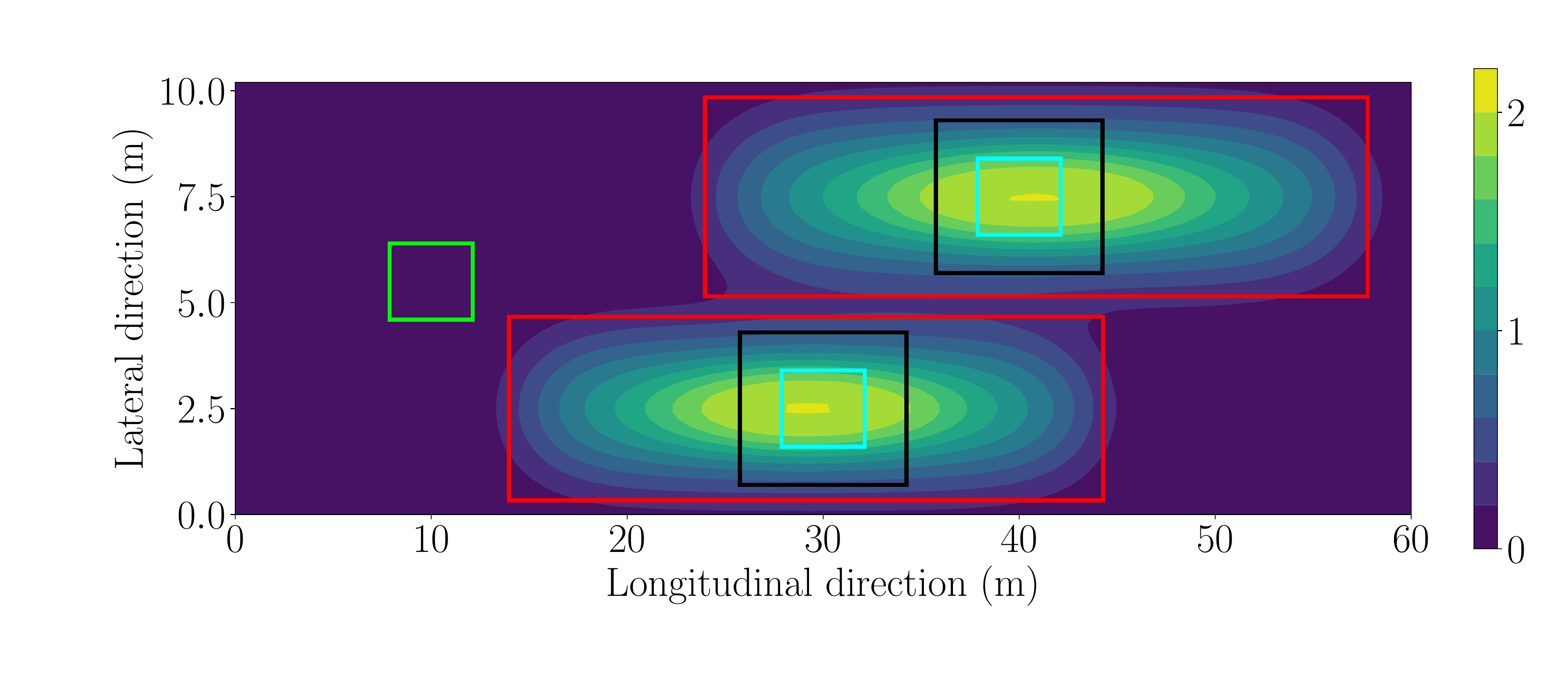}
%\vspace{-10mm}
\caption{Illustration of obstacle avoidance function}\label{fig_col}
%\vspace{-8mm}
\end{figure}

Specifically, the first $\left\{1-\tanh\left[\left(\frac{x_1 - \delta_{oi}}{0.5d_1}\right)^{p_1}+\left(\frac{x_2 - o_{i2}}{0.5d_2}\right)^{p_2}\right]\right\}$, is designed such that its contours have a nearly flat lateral edge and a curvy, sliding-prone longitudinal edge (see \Cref{fig_col}), in accordance with requirement (i). This contour shape helps in accommodating more vehicles on a given road width, while also allowing the vehicles to overtake, whenever possible, thanks to the curvy shape and gradient on the longitudinal edge. This is achieved by choosing a higher value of $p_1$ compared to $p_2$. The drawback of this design (by itself) is that it does not fulfil requirement (ii), as it creates a flat area over the top of the ellipsoid, with almost zero gradient, which affects the convergence rate of the numerical solution algorithm. This drawback may be repaired by adding a second ellipsoid, $\left\{\left[\left(\frac{x_1 - \delta_{oi}}{0.25d_1}\right)^{p_3}+\left(\frac{x_2 - o_{i2}}{0.25d_2}\right)^{p_4}\right]^{p_5} + 1 \right\}^{-1}$, which barely alters the opportune contours of the first ellipsoid around the obstacle; while, with appropriate values of its exponents, it creates a strong gradient of the potential function at all points up until the centre of the ellipsoid, so that requirement (ii) is satisfied as well. 

In summary, the sub-objective function in presence of   obstacles is $\sum_{i=1}^{n}\left[c_i\left(x,\ o_i\right)\right]$. 

To illustrate the design, contour plots of the sub-objective \cref{ocp-eq3} for two obstacles, as seen by an EV, are shown in \Cref{fig_col}. All the three vehicles have same dimensions of 4.25 m length and 1.8 m width. The EV is positioned at (10, 5.5) m on a road of width 10.2 m, as shown in \Cref{fig_col} with a green box. The longitudinal and lateral speeds of the EV are 30 m/s and 0.75 m/s, respectively. Two obstacles 1 and 2 are positioned at (30, 2.5) m and (40, 7.5) m with speeds (25, 0) m/s and (35, 0) m/s, respectively, and are marked with cyan boxes. The combined dimensions of both ego and corresponding obstacle vehicles are represented by black boxes. The time-gaps $\left(\omega_1,\ \omega_2\right)$ are (0.35, 0.5) s. Red boxes represent the major and minor axes lengths $d_1$ and $d_2$ of \cref{ocp-eq3}. The safety coefficients $\mu_x$ and $\mu_y$ are 1.3 and 1.2, respectively. Since the longitudinal speeds are different for each of the vehicles, the position of the centre of the longitudinal axis of the ellipsoid is shifted towards the EV for obstacle 1; and away from the EV for obstacle 2, according to \cref{ocp-eq1}. It can also be observed that the width of the red box of obstacle 2 is slightly larger that of obstacle 1, which is because the EV is approaching obstacle 2, in accordance with \cref{ocp-eq2}. The values of the exponents are $p_1=6,p_2=p_3=p_4=p_5=2$. It can be observed that the contours are flat at the lateral edge, facilitating better usage of lateral space; while featuring strong curvature on the longitudinal edge, which facilitates smooth overtaking by a faster EV. Furthermore, the displayed subsequent-contour distances indicate a strong gradient up until the ellipsoid centre, where a maximum value of function $c_i$ is reached. Finally, it may be seen that the ellipsoid extends both upstream and downstream of each obstacle, hence it may cause both repulsing and nudging influence on the EV.

\subsubsection{Coupling of Longitudinal and Lateral Speeds}\label{long_cpl}
The longitudinal and lateral movements are assumed to be decoupled, according to the vehicle dynamics in \cref{eq1}, which is largely a valid assumption in medium and high longitudinal speeds without strong turning angles. In the case of very low longitudinal speeds, however, there could be some infeasible lateral movements, which would necessitate very high steering value for the real vehicle. Such situations correspond to a high value of the rate of lateral speed over longitudinal speed. To suppress such situations, a sub-objective that penalizes the deviation of that rate, beyond a certain small parameter $\beta$, as a quadratic function, is given as

\begin{align}\label{ocp-eq4}
f_c=\begin{cases}
\left[\beta x_3(k)-|x_4(k)|\right]^2 & \text{if } |x_4(k)|>\beta x_3(k) \\
0 & \text{otherwise}
\end{cases}
\end{align}
The sub-objective in \cref{ocp-eq4} acts as an indirect coupling between longitudinal and lateral speeds. It is noted that this helps in avoiding the usage of complex nonlinear vehicle dynamics, which would be less efficient at obtaining a numerical solution within a very small computation time. 
 
\subsubsection{Deviation of Longitudinal Acceleration Between Planning Horizons}
Each finite-time planning horizon, within the MPC scheme, can start with any feasible longitudinal acceleration. This may result in a strong change of the acceleration between the last applied value $u_{1\mathrm{prev}}$ and the first value $u_1\left(0\right)$ of the new planning horizons. This can lead to discomfort for the passengers and can be mitigated by penalizing the deviation of the starting acceleration $u_1\left(0\right)$) of a planning horizon from the last applied value of acceleration of the previous planning horizon $u_{1\mathrm{prev}}$ , i.e.,

\begin{align}\label{ocp-eq5}
f_d=\left[u_1(0)-u_{1\text{prev}}\right]^2
\end{align}
\subsection{OCP Formulation and Necessary Conditions for Optimality}
The OCP is defined as the minimization of the objective function, which is a combination of all the sub-objectives discussed above, subject to the equality constraints (state equations) in \cref{eq1} and the inequality constraints (bounds on control) in \cref{sdbc-eq13}. 

Summarizing \Cref{obj}, we have the objective function
\begin{align}
J= & \sum_{k=0}^{K-1} \big\{w_1 \left(u_1(k)\right)^2 + w_2 \left(u_2(k)\right)^2 + w_3 \left(x_3(k) - v_{d_1}\right)^2 \nonumber \\ & + w_4 (x_4(k)-v_{d_2})^2 + w_5 \sum_{i=1}^{n}[c_i(\boldsymbol{x},\boldsymbol{o}_i)] + w_6 f_c\big\} + w_7 f_d\label{ocp1}
\end{align}
where $K$ is the planning horizon; and $w_1$ to $w_7$ are weighting factors of each of the sub-objectives to be chosen appropriately, so as to reflect the relative importance and typical range of values of each sub-objective. More specifically, the values of the weighting factors are decided via a manual trial-and-error procedure. Using some initial weights, the vehicle's behaviour is observed and evaluated in preliminary simulations involving multiple driving episodes; weights are increased or lowered if corresponding sub-objectives are deemed to over-perform (to the detriment of other sub-objectives) or to under-perform; and this is continued, until an acceptable overall vehicle behaviour has been reached. Once appropriate weights have been specified in this way, the sensitivity of optimal control decisions with respect to different weight values of the same order is usually low. 

It should be noted that the use of an extremely high weight for a sub-objective may lead to an accordingly ill-conditioned optimisation problem that can hardly be solved via gradient-based numerical solution algorithms \cite{fletcher1987}. In the present application, the most important sub-objective is the one referring to collision avoidance. However, if the related weight $w_5$ is given a very high value, so as to guarantee that a collision will literally never occur in the produced optimal vehicle trajectory under any circumstances, then the resulting OCP may be very difficult to solve numerically. Instead, one may select a reasonably high value for $w_5$, which does not hinder the numerical solution, but may, in very rare circumstances, result in a collision trajectory. This is the approach pursued in this work. A possible produced collision trajectory may be readily identified, in which case an emergency situation is declared, and the OCP is reformulated with stricter constraints to produce a collision-free vehicle trajectory according to \Cref{mpc}.

The general form of the objective function, which is used in the further analysis, is given by

\begin{equation}\label{ocp2}
J = \sum_{k=0}^{K-1} \Phi[\boldsymbol{x}(k),\boldsymbol{u}(k)]
\end{equation}
\noindent
where $u$ is the control vector, comprising the longitudinal and lateral accelerations. The general vector-based representation of \cref{eq1} is 
\begin{equation}\label{ocp3}
\boldsymbol{x}(k+1)=\boldsymbol{f}[\boldsymbol{x}(k),\boldsymbol{u}(k)]
\end{equation}
\noindent
The Hamiltonian function, including the equality \cref{ocp3} and inequality \cref{sdbc-eq13} constraints, is defined as
\begin{align}\label{ocp4}
H[\boldsymbol{x}(k),\boldsymbol{u}(k), \boldsymbol{\lambda}(k+1), \boldsymbol{\mu}(k)] = \Phi[\boldsymbol{x}(k),\boldsymbol{u}(k)] + \nonumber \\ \boldsymbol{\lambda}(k+1)^T \boldsymbol{f}[\boldsymbol{x}(k),\boldsymbol{u}(k)] + \boldsymbol{\mu}(k)^T \boldsymbol{h}[\boldsymbol{x}(k),\boldsymbol{u}(k)]
\end{align}
\noindent
where $\boldsymbol{\lambda}(k)$ are the co-states, associated with the state equations, and $\boldsymbol{\mu}(k)$ are multipliers, associated with the control constraints. Based on these, the necessary conditions of (local) optimality, to be used in the numerical solution algorithm, are given next. We have the state equation

\begin{align}\label{ocp5}
\boldsymbol{x}(k+1) = \frac{\partial H}{\partial \boldsymbol{\lambda}(k+1)} = \boldsymbol{f}[\boldsymbol{x}(k),\boldsymbol{u}(k)].
\end{align}
\noindent
The control condition is given by
\begin{align}\label{ocp6}
\partial H/\partial \boldsymbol{u}(k) = \boldsymbol{0}.
\end{align}
\noindent
The co-state equation is given by
\begin{align}\label{ocp7}
\boldsymbol{\lambda}(k) = \partial H/\partial \boldsymbol{x}(k).
\end{align}
\noindent
The complementarity conditions are
\begin{align}\label{ocp8}
\mu_i(k)h_i[\boldsymbol{x}(k),\boldsymbol{u}(k)]=0, \;\mu_i(k) \geq 0
\end{align}
\noindent
Finally, the boundary conditions are given by
\begin{subequations}\label{ocp9}
\begin{align}
\boldsymbol{x}(0)&=\boldsymbol{x_0}\label{ocp9a}\\
\boldsymbol{\lambda}(K)&=\boldsymbol{0}\label{ocp9b}.
\end{align}
\end{subequations}

\subsection{Numerical Solution Algorithm}\label{num_sol}
The solution of the described OCP must be obtained through an efficient numerical solver that enables real-time feasibility. A very efficient gradient-based feasible direction algorithm (FDA) \cite{fda,papageorgiou2015} is employed to solve the present OCP. The underlying method leverages the explicit structure of the state equations and uses the necessary conditions of optimality to map the OCP into a Nonlinear Programming (NLP) problem in the reduced space of control variables. Thus, the algorithm attempts to reach a control trajectory $\boldsymbol{u}(k), k = 0,\ldots , K - 1$, which corresponds to a local minimum of the cost function in the $mK$-dimensional space, where $m$ is the number of control variables. This marks a substantial reduction of the problem dimension, as the state variables are eliminated. 

More specifically, FDA exploits the fact that $\boldsymbol{g}(k)=\left[\partial \boldsymbol{f}/\partial \boldsymbol{u}(k)\right]^\text{T} \boldsymbol{\lambda}(k+1) + \partial \Phi / \partial \boldsymbol{u}(k)$ equals the reduced gradient in the $mK$-dimensional reduced space of the control, if the states and co-states involved in this partial derivative satisfy the state and co-state equations. Having this possibility to calculate reduced gradients, FDA is an iterative procedure, starting with a feasible initial-guess control trajectory specified by the user. The algorithm can be readily extended to consider upper and lower (possibly state-dependent) bounds on the control variables \cite{fda,papageorgiou2015}. 

Each FDA iteration uses the updated gradient $g\left(k\right)$ to compute an appropriate descent direction (e.g., conjugate gradient or quasi-Newton direction) and produce, based on line search, an improved feasible control trajectory that reduces the objective function value, compared to the previous iteration, while satisfying the state equations and control constraints. The improved control trajectory is the starting point of the next iteration, and so forth. The algorithm features global convergence, i.e., it reaches a local minimum starting from any feasible initial-guess trajectory \cite{fletcher1987}. The minimum is practically reached, and the algorithm's iteration are stopped, when the magnitude of the gradient approaches sufficiently a zero value. All necessary conditions of optimality are satisfied at the algorithm;s convergence. It should be noted that an initial-guess trajectory that is closer to the optimal one may reduce the required number of iterations. 

Among several conjugate-gradient and Quasi-Newton methods, the Polak-Ribiere method was found most efficient for this OCP and is used for calculating search directions inside FDA. The approach is fast enough to be considered for real-time applications. With appropriate tuning of some algorithm parameters, the algorithm's runtime to generate vehicle trajectories, for a planning horizon of 8 s with a time-step size of $T=250$ ms (i.e., for $K=32$), is 12.4 ms on average (less than 100 ms in 99.9\% plans, less than 250 ms in 99.99\% plans) on a machine powered by Intel Core i5-8500 CPU operating at a maximum of 3.00 GHz frequency. The maximum computation time observed is 1.71 s and the minimum of 0.3 ms. It should also be noted that FDA iterations may be stopped at any time, even before convergence, delivering a control trajectory that may not be the optimal one, but is feasible, i.e., it satisfies all state equations and control constraints. 

In the OCP at hand, all equality and inequality constraints are linear, and all sub-criteria are convex with one exception, namely the collision avoidance term, which is non-convex. Thus, due to the collision avoidance term, the OCP as a whole is non-convex, and this implies that FDA may converge to a local minimum that satisfies all necessary conditions of optimality. Although it is not possible to know, for any delivered minimum, whether it is a local or global one, we never identified, in the extensive simulation investigations of \Cref{sim}, any awkward EV manoeuvres that might correspond to a bad local minimum. 

\begin{figure*}
\centering
\begin{subfigure}{0.4\columnwidth}
\centering
\includegraphics[scale=1,width=\columnwidth]{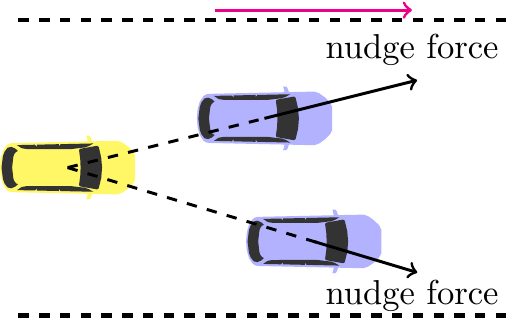}
\caption{}\label{fig_nudge_a}
\end{subfigure}\hspace{1cm}
\begin{subfigure}{0.4\columnwidth}
\centering
\includegraphics[scale=1,width=\columnwidth]{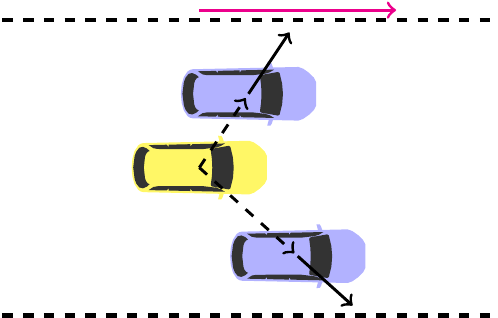}
\caption{}\label{fig_nudge_b}
\end{subfigure}\hspace{1cm}
\begin{subfigure}{0.4\columnwidth}
\centering
\includegraphics[scale=1,width=\columnwidth]{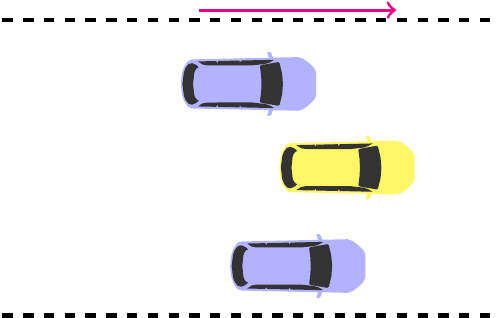}
\caption{}\label{fig_nudge_c}
\end{subfigure}
\caption{A microscopic effect of vehicle nudging: The yellow vehicle nudges the slower blue vehicles aside to pass.}\label{fig_nudge}
\end{figure*}

One of the reasons why bad local optima are not encountered is the strongly dynamic nature of traffic on a highway, as considered here. Another reason is the utilised vehicle nudging. It is known from moving robotics applications \cite{hwang1992} that the potential-function like terms, as those used here for collision avoidance, may create a local minimum on a trajectory, e.g., according to \Cref{fig_nudge_a}, where a fast car is trapped behind two slower vehicles that drive in parallel symmetry. If no vehicle nudging would be present, the car would follow the slow vehicles, taking a locally optimal trajectory. On the other hand, if nudging is present, as in our OCP, the car may nudge both vehicles aside (\Cref{fig_nudge_b}) and pass (\Cref{fig_nudge_c}). Many such microscopic episodes were identified in the simulations reported in \Cref{sim} and they illustrate that vehicle nudging is, in several situations, factually suppressing bad local-minimum trajectories.

\subsection{Model Predictive Control (MPC)}\label{mpc}
MPC has been established as a quasi-feedback control scheme that enables the deployment of optimisation and optimal control methods in real time \cite{mayne2014}. The present application is labelled as economic MPC \cite{ellis2014}, because it aims at minimising a quasi-physical objective criterion, rather than tracking desired values for system states.   

In the present application, FDA is embedded in an MPC scheme to deliver in real time numerical solutions of the formulated OCP that determine appropriate vehicle trajectories. In order to investigate and assess the approach under the novel requirements of lane-free traffic and vehicle nudging, as well as to evaluate its impact at traffic level, the MPC scheme is applied, in the next section, independently to all vehicles populating a ring-road infrastructure. The OCP looks into the future, according to its time-horizon, thus avoiding myopic actions by anticipating future action needs. This is deemed beneficial for the pursued vehicle tasks, specifically for higher efficiency (vehicle advancement), safety (collision avoidance) and passenger convenience (smooth trajectories). On the other hand, due to the dynamic nature of traffic and the inherent uncertainty about the movement of other vehicles around an EV, the consideration of very long time horizons for the OCP is not sensible, as the prediction of movement of other vehicles becomes increasingly inaccurate with time. It was found empirically that time horizons in the order of 8 s are suitable in terms of performance, as well as to account for the dynamic, but also uncertain, nature of traffic. 
 
MPC calls the FDA in an event-based manner for each vehicle, and this results in asynchronous trajectory computation and application in emulated real-time for each of the vehicles. Each EV receives the current states and the latest computed trajectories of all other vehicles (obstacles) within its IZ. Since the OCP is solved at different instances for each of the vehicles, each EV usually holds, at the time it computes its next optimal trajectory, only partial decided trajectories of the obstacles. These partial trajectories are extrapolated, to cover the whole planning horizon of the EV, by considering zero accelerations (both longitudinally and laterally) beyond the known obstacle trajectory parts; and the completed obstacle trajectories are used in the obstacle avoidance sub-objective of OCP (\Cref{obst_avoid}) and in the emergency constraints (\Cref{sdbc}) within the OCP. 

Each FDA call is initiated by the MPC framework in an event-based mode, each time with updated initial states of the EV and updated obstacle trajectory predictions. The following events trigger the FDA calls, i.e., the computation of a new trajectory for the EV:

\begin{itemize}
\item The vehicle has already applied the part of the last generated trajectory, which corresponds to a pre-specified \textit{application period}, which is here half of the planning horizon, and the reason why the planning horizon is larger than the application period is in order to prevent the vehicle from taking myopic actions. 
\item There is a substantial deviation (0.2 m longitudinally or 0.1 m laterally) in the trajectory of any of the dynamic obstacles. This is necessary to prevent collisions due to inaccurate consideration of the actual obstacles' movement.
\item A new obstacle enters the IZ of the EV.
\end{itemize}

In all the above events, the unused control trajectory part is used as a good initial guess for FDA, so as to obtain faster convergence. The rest of the guess trajectory, beyond the unused trajectory part, is assumed to have zero accelerations. The trajectories generated at each of the iterations are always feasible by construction, hence the FDA may be stopped at any time, especially in the very rare case it is taking a lot of iterations. 

\begin{figure*}
\centering
\begin{subfigure}{0.4\columnwidth}
\centering
\includegraphics[scale=1,width=\columnwidth]{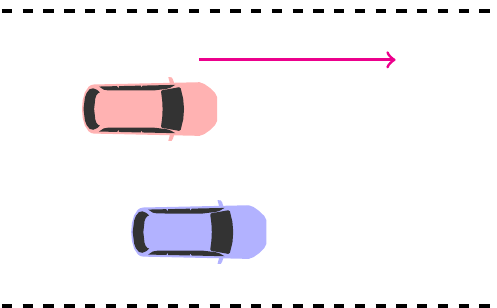}
\caption{}\label{fig_crash_a}
\end{subfigure}\hspace{1cm}
\begin{subfigure}{0.4\columnwidth}
\centering
\includegraphics[scale=1,width=\columnwidth]{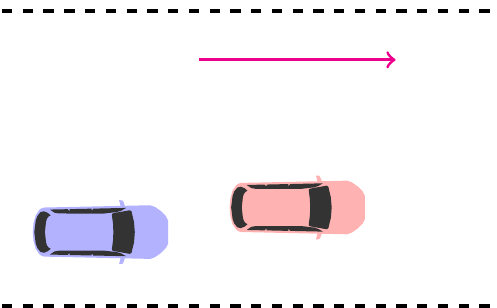}
\caption{}\label{fig_crash_b}
\end{subfigure}\hspace{1cm}
\begin{subfigure}{0.4\columnwidth}
\centering
\includegraphics[scale=1,width=\columnwidth]{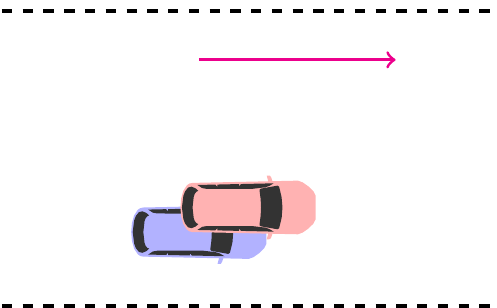}
\caption{}\label{fig_crash_c}
\end{subfigure}
\caption{Illustration of collision detection: \subref{fig_crash_a} Longitudinal overlap; \subref{fig_crash_b} Lateral overlap; \subref{fig_crash_c} Longitudinal and lateral overlap (collision).}\label{fig_crash}
\end{figure*}

As explained earlier, minimization of the objective function does not guarantee generation of a collision-free trajectory for the EV in all possible circumstances. Therefore, the trajectories generated by the FDA are checked for possible collisions, based on the predicted trajectories of the obstacles. The checks are performed for possible longitudinal or lateral collision, and, if such a collision is indeed detected, a respective new OCP is specified, including appropriate preventive measures. %The details of the two checks to detect possible longitudinal or lateral collisions, respectively, with any obstacle may be found in \cite{karteek2022}.
A longitudinal collision with $i^\text{th}$ obstacle is detected if all the following checks return true:

\begin{itemize}
\item $x_1\left(0\right)<o_{i1}\left(0\right)$, i.e., initial longitudinal position of EV's centre is behind the obstacle's centre, otherwise no longitudinal collision is conceivable.
\item $\displaystyle \left(o_{i1}\left(k\right)-\frac{e_l}{2}-\frac{o_{il}}{2}-\omega_{1em}x_3\left(0\right)\right)<x_1\left(k\right)<\left(o_{i1}\left(k\right)+\frac{e_l}{2}+\frac{o_{il}}{2}+\omega_{1em}x_3\left(0\right)\right)$, for any $k$ within the time horizon $0 \le k < K$, to check longitudinal overlap of length and emergency time-gap on rear and front sides of the EV, where $\omega_{1em}=0.5\omega_1$. \Cref{fig_crash_a} depicts a case of longitudinal overlap.
\item $\displaystyle \left(o_{i2}\left(k\right)-\frac{e_w}{2}-\frac{o_{iw}}{2}-\epsilon\right)<x_2\left(k\right)<\left(o_{i2}\left(k\right)+\frac{e_w}{2}+\frac{o_{iw}}{2}+\epsilon\right)$, for any $k$ within the time horizon $0\le k\ <\ K$, to check lateral overlap of width on either side of the EV, with a small positive safety margin $\epsilon$. \Cref{fig_crash_b} depicts a case of lateral overlap.
\end{itemize}

\Cref{fig_crash_c} depicts a case of both longitudinal and lateral overlaps, which is a collision. If the initial position of the EV is behind the obstacle and overlaps are detected both longitudinally and laterally in the trajectories, this is categorized as a longitudinal collision. Note that, due to the presence of $\omega_{1em}>0$ and $\epsilon>0$, the detected collision may not be a physical one. Also, if the second and third conditions above are satisfied for different times $k$, no physical collision may be present. Nevertheless, such situations may reflect a risky vehicle interference that should better be avoided.

Similarly to the longitudinal collision detection, a lateral collision is detected if all the following checks return true:

\begin{itemize}
\item $\displaystyle \left(o_{i1}(0)-\frac{e_l}{2}-\frac{o_{il}}{2} - \epsilon \right) < x_1(0)<\left(o_{i1}(0)+\frac{e_l}{2}+\frac{o_{il}}{2} + \epsilon \right)$, to check the lateral alignment at the start of trajectory. 
\item $\displaystyle \left(o_{i1}(k)-\frac{e_l}{2}-\frac{o_{il}}{2} - \epsilon \right) < x_1(k) < \left(o_{i1}(k)+\frac{e_l}{2}+\frac{o_{il}}{2} + \epsilon \right)$, for any $k$ within the time horizon $0\leq k < K$, to check the overlap longitudinally.
\item $\displaystyle \left(o_{i2}(k))-\frac{e_w}{2}-\frac{o_{iw}}{2} - \epsilon \right) < x_2(k)< \left(o_{i2}(k))+\frac{e_w}{2}+\frac{o_{iw}}{2} + \epsilon \right)$, for any $k$ within the time horizon $0\leq k < K$, to check the overlap laterally.
\end{itemize}
Thus, if the initial positioning of the vehicles is as shown in \Cref{fig_crash_a} and evolves to a position as shown in \Cref{fig_crash_c} during the course of the trajectory, this is categorized as a lateral collision.

It must be noted that the two checks are non-exclusive mutually, hence the outcome may indicate both a longitudinal and a lateral collision. In this case, the lateral collision characterisation prevails, since both vehicles are certainly laterally aligned at the starting time ($k=0$) due to the first bullet above. 

When a longitudinal collision is detected with the $i^{\text{th}}$  obstacle, the OCP is re-formulated to include a new state-dependent bound on the upper limit of longitudinal acceleration, as described in \Cref{sdbc-1}1, so that the EV is instructed to follow the $i^{\text{th}}$  obstacle, with which it was detected to interfere during the first run. When a lateral collision is detected, the OCP is reformulated to include new state-dependent upper and lower bounds on lateral acceleration, as described in \Cref{sdbc-2}. The new bounds restrict the lateral movement of the EV within $\pm0.15$ m of the initial lateral position at the start of the respective planning horizon. This is achieved by assuming two virtual boundaries at $x_2\left(0\right)\pm0.15$ m, parallel to the actual road boundaries, but with a very narrow corridor for lateral movement, which effectively prevents the lateral collision observed in the first run.

In the investigations of \Cref{sim}, myriads of trajectory planning runs have been executed. Collisions in the optimal trajectory were detected only for a very tiny portion of them (see \Cref{tb2}). The triggered prevention measures in those cases were always successful, i.e., there was no single case where the second, more strongly restricted trajectory planning would deliver a collision trajectory.

Each vehicle has a pre-defined constant longitudinal desired speed $V_{{des}_1}$, and it is ideally desired that it moves close at that speed, possibly by manoeuvring around encountered slower obstacles, whenever possible. However, considering $V_{{des}_1}$ directly in the desired-speed term of OCP's objective function (\Cref{des_sp}) may result in unreasonably high values for longitudinal acceleration if the initial longitudinal speed $x_3(0)$ is low, because, in such a case, the sub-objective for desired speed would assume accordingly high values that the optimisation would strive to reduce by increasing the longitudinal acceleration, something that may affect passenger convenience. 

A similar issue arises in cases of high vehicle density, where there is very little space for lateral movement and overtaking of downstream obstacles. In such conditions, vehicles tend to move in clusters at similar longitudinal speeds, that are clearly lower than everybody's desired speed, and with only small relative displacements. Similar conditions are encountered in conventional lane-based traffic, when speed in all lanes is low, due to high traffic density (congestion), and there is hardly any escape forward, even for fast vehicles. Maintaining a high desired speed in the OCP in these circumstances, may result in an aggressive forward-acceleration behaviour of the EV, which is pointless, since overtaking is hardly possible. Such aggressive behaviour may lead to collisions, but may also cause acceleration bursts and hence discomfort to the passengers. 

To address both these issues, an adaptive desired speed $v_{d_1}$ is defined in \cref{ocp9}, which is computed prior to every new trajectory calculation for use in the desired-speed sub-objective of \Cref{des_sp}.  Based on \cref{ocp9}, $v_{d_1}$ may be limited to values lower than $V_{\text{des}_1}$, to address the above two concerns. Specifically, the adaptive desired speed $v_{d_1}$ is always limited to a positive increment $V_{{\text{incr}}_1}$ above the current longitudinal ego speed $x_3(0)$, which addresses the first issue above, since the desired speed targeted by the OCP is never higher than $V_{{\text{incr}}_1}$ above the current vehicle speed $x_3\left(0\right)$, hence the longitudinal acceleration will not take unreasonably high values. 

To address the second issue above, a second limit is included in (29), which is considered only at dense traffic, i.e., if $D_d>D_{\text{bar}}$ , where $D_d$ (in veh/km) is the downstream vehicle density, defined as the number of obstacles included in the downstream part of the IZ, divided by the zone's length; and $D_{\text{bar}}$  is a density threshold, selected around the critical density value. Specifically, if $D_d>D_{\text{bar}}$ , the adaptive desired speed $v_{d_1}$ is additionally limited to a positive increment $V_{{\text{incr}}_2}$ above the current average downstream longitudinal speed $D_v$ of the vehicles included in the downstream part of the IZ. This second limit addresses the second issue above, as, at higher densities, the EV has no strong incentive to accelerate beyond the average speed of surrounding vehicles.

Of course, it may be seen in \cref{ocp9}, that the adaptive desired speed $v_{d_1}$ may equal the actual desired speed  $V_{{des}_1}$, e.g., at lower densities and high initial speed $x_3(0)$.

\begin{align}\label{ocp9}
v_{d_1}= 
\begin{cases}
\text{min}\left\{x_3(0) + V_{\text{incr}_1}, V_{\text{des}}\right\} & \text{if}\; D_d \leq D_\text{bar}\\
\text{min}\left\{x_3(0) + V_{\text{incr}_1}, D_v + V_{\text{incr}_2}, V_{\text{des}}\right\} & \text{if}\;  D_d > D_\text{bar}
\end{cases}
\end{align}

\section{Simulation Results}\label{sim}
\subsection{Simulation Environment and Scenarios}
To run traffic-level simulations in a lane-free environment, a custom-made extension, called TrafficFluid-Sim \cite{dimitri2021-2}, was developed within the well-known and broadly used Simulation of Urban Mobility (SUMO) simulator \cite{sumo}. The open-source codebase of SUMO is extended and used to communicate control inputs and other information with SUMO. The MPC framework is incorporated into the API and invokes the FDA algorithm when needed. Due to the lane-free structure of the road, the default car-following and lane-changing models of SUMO are not appropriate for the vehicle movement; therefore, the extension is utilized to incorporate the proposed vehicle movement strategies. SUMO detectors, measuring flow, are used, but, due to lane-free traffic, they are set to span over the whole road width, at their specific locations, thus measuring the total cross-section flow.

The described vehicle-movement strategy is based on the formulated OCP, its numerical solution via FDA and its MPC deployment in real time. To test the strategy's vehicle-level properties, as well as to assess the emerging traffic-level characteristics, the strategy is applied independently to all vehicles driving on an unfolded lane-free ring-road of 1 km length and 10.2 m width. Distinct simulations are carried out for corresponding vehicle densities. Eight classes of vehicles are considered at random, the corresponding dimensions of each class being given in \cref{tb1}. At each simulation, vehicles start with zero speed.
 %and roughly homogeneous initial placement within the 2-d ring-road surface, see \cite{karteek2022} for details. Vehicles are assigned longitudinal desired speeds from the range [25, 35] m/s.

For homogeneous initial placement of the vehicles with appropriate two-dimensional spacing between each other, the lateral space is divided into four virtual lanes, and the longitudinal space is divided into $n/4$ sections, where $n$ is the number of vehicles (density) being considered. This creates exactly $n$ cells, homogeneously distributed on the two-dimensional ring-road surface, and each vehicle is placed around the centre of a cell, with some two-dimensional random deviations. The assignment of longitudinal desired speeds to individual vehicles is based on their initial lateral positioning. Specifically, the considered longitudinal desired speed range is [25, 35] m/s, and this range is divided into four equal desired speed zones, one for each virtual lane. Each vehicle has its own longitudinal desired speed, assigned randomly with a uniform distribution within its speed zone, based on initial lateral position. Note that, these desired speed zones are for initial placement only, to aid the movement in high densities. Lateral desired speeds are zero for all vehicles.

\begin{table}[!h]
\centering
\caption{Vehicle classes and dimensions.}\label{tb1}
\begin{tabular}{lccccccccc}
%\begin{tabular}{|l|
%>{\centering\arraybackslash}p{1.75em}|
%>{\centering\arraybackslash}p{2.25em}|
%>{\centering\arraybackslash}p{2.25em}|
%>{\centering\arraybackslash}p{2.25em}|
%>{\centering\arraybackslash}p{2.25em}|
%>{\centering\arraybackslash}p{2.25em}|
%>{\centering\arraybackslash}p{2.25em}|
%>{\centering\arraybackslash}p{2.25em}|
%>{\centering\arraybackslash}p{1.75em}|
%>{\centering\arraybackslash}p{1.75em}|
%>{\centering\arraybackslash}p{1.75em}|}
%\begin{tabularx}{0.48\textwidth} { 
%   >{\raggedright\arraybackslash}X 
%   >{\centering\arraybackslash}X 
%   >{\centering\arraybackslash}X 
%   >{\centering\arraybackslash}X 
%   >{\centering\arraybackslash}X 
%   >{\centering\arraybackslash}X 
%   >{\centering\arraybackslash}X }
\toprule
Class & I & II & III & VI & V  & VI & VII & VIII \\ 
\midrule
Length (m) & 3.2 & 3.4 & 3.9 & 4.25 & 4.55 & 4.6 &5.15 & 5.2\\
\midrule
Width (m) & 1.6	 & 1.7 & 1.7 & 1.8 & 1.82 & 1.77 & 1.84 & 1.88 \\ 
\bottomrule
\end{tabular}
\end{table}

After all vehicles are placed and are assigned their longitudinal desired speeds, they start moving, from zero initial speed, according to the developed strategy, and they strive to reach their desired speed, while avoiding other vehicles in front and around them.
 
The following parameters are used in the simulations: In \cref{ocp1}, $\{w_1,w_2,w_3,w_4,w_5,w_6,w_7\} = \{0.005, 0.005, 0.015, 0.005, 7.0, 0.1, 0.005\}$; longitudinal safety parameter $\omega_1=0.53$ s and lateral safety parameter $\omega_2=0.5$ s; $\epsilon_w = 0.1$ in \cref{ocp-eq1,ocp-eq2}; in \cref{ocp-eq3}, $p_1=6,\;p_2=p_3=p_4=p_5=2$; in \cref{ocp-eq4}, $\beta=0.03$; the length of each part of the IZ is not allowed to fall below 100 m; density threshold $D_\text{bar}=150$ veh/km in \cref{ocp9}; discrete-time step $T=250$ ms; time-horizon 8 s, hence $K=32$; constant bounds on acceleration $U_\text{max1}=0.5\; \text{m}/\text{s}^2$, $U_\text{min1}=-2\; \text{m}/\text{s}^2$ for regular case and $U_\text{min1}=-4\; \text{m}/\text{s}^2$ for emergency case.

\subsection{Traffic-level Results}
\begin{figure}[!t]
\centering
%\vspace{2mm}
\includegraphics[scale=0.219]{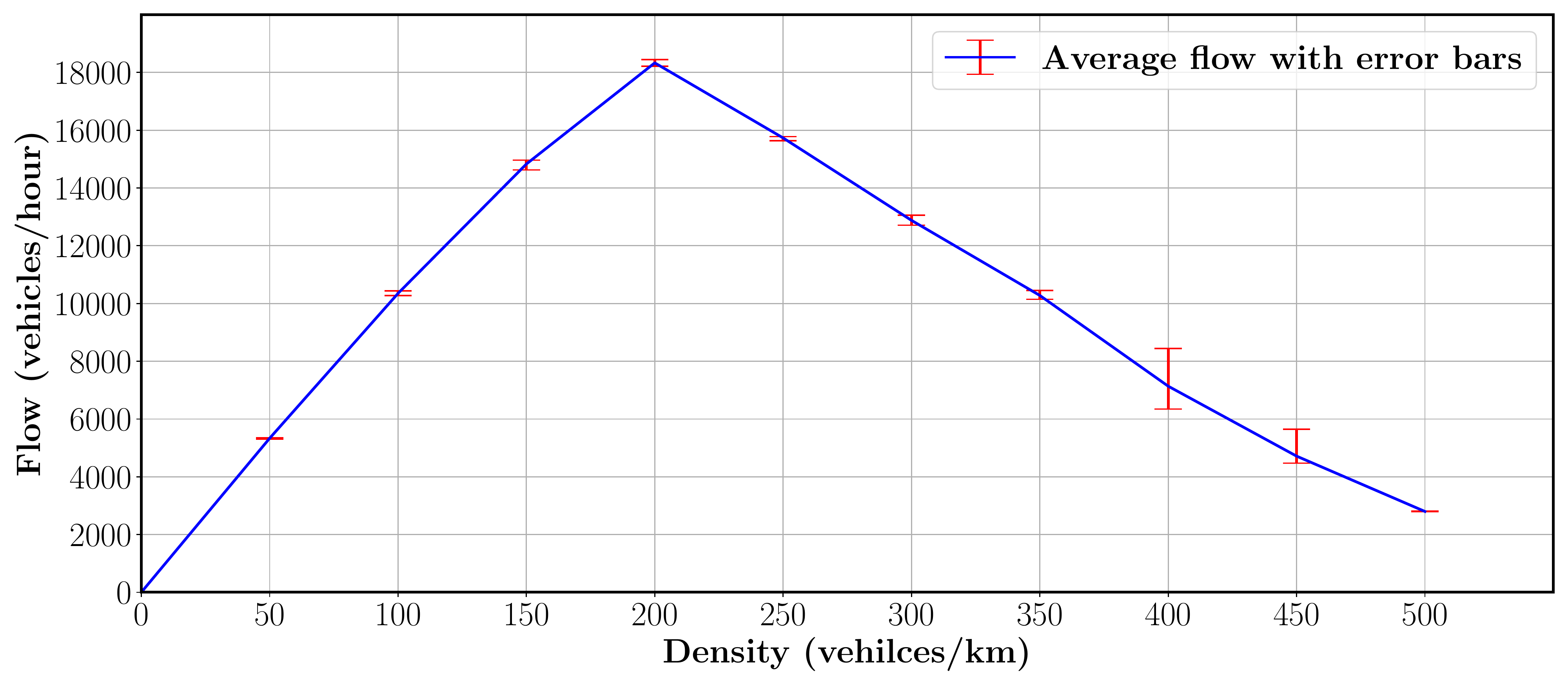}
\caption{Fundamental diagram}\label{fig_fd}
%\vspace{-18pt}
\end{figure}

Simulations are run for densities ranging from 50 to 500 veh/km. For each simulation with a specific density, a corresponding flow value is obtained from loop detectors. Specifically, five loop detectors are placed at distances of 200 m from each other, and the overall flow values are obtained by taking the average of the flow values recorded at each of the loop detectors. For each specific density value, five different random initializations (different seed values) of desired speeds, within the specified range, and of starting positions are considered in corresponding simulation replications. The average flows (over the five replications) are displayed against the corresponding density values to form a fundamental diagram (FD), with error bars reflecting maximum and minimum deviations among the five replications, as provided in \cref{tb2} and \cref{fig_fd}.

\begin{table*}
\centering
\caption{Average flow outcomes (5 replications) for given densities.}\label{tb2}
\begin{tabular}{lccccccccccc}
%\begin{tabular}{|l|
%>{\centering\arraybackslash}p{1.75em}|
%>{\centering\arraybackslash}p{2.25em}|
%>{\centering\arraybackslash}p{2.25em}|
%>{\centering\arraybackslash}p{2.25em}|
%>{\centering\arraybackslash}p{2.25em}|
%>{\centering\arraybackslash}p{2.25em}|
%>{\centering\arraybackslash}p{2.25em}|
%>{\centering\arraybackslash}p{2.25em}|
%>{\centering\arraybackslash}p{1.75em}|
%>{\centering\arraybackslash}p{1.75em}|
%>{\centering\arraybackslash}p{1.75em}|}
%\begin{tabularx}{0.48\textwidth} { 
%   >{\raggedright\arraybackslash}X 
%   >{\centering\arraybackslash}X 
%   >{\centering\arraybackslash}X 
%   >{\centering\arraybackslash}X 
%   >{\centering\arraybackslash}X 
%   >{\centering\arraybackslash}X 
%   >{\centering\arraybackslash}X }
\toprule
Density (veh/km) & 50 & 100 & 150 & 200 & 250  & 300 & 350 & 400 & 450 & 500\\ 
\midrule
Flow (veh/h) & 5334 & 10350 & 14830 & 18332 & 15733 & 12879 & 10278 & 7137 & 4715 & 2799 \\
\midrule
Emergency \%  & 0.0 & 0.001 & 0.002 & 0.0 & 0.0 & 0.0 & 0.001 & 0.030 & 0.0 & 0.0143\\ 
\bottomrule
\end{tabular}
\end{table*}

A first, expected, qualitative observation is that the emerging FD has a similar inverse-U (or, rather, inverse-V) shape, as in conventional lane-based traffic. The left-hand (under-critical) side of the FD approaches a straight line, particularly at lower densities, with a slight bending downwards when the critical density (200 veh/km) is approached.  The slope of this line is around 28 m/s, which is slightly lower than the average desired speed of all vehicles. This implies that, for under-critical densities, the vehicles, guided by the developed movement strategy, find ways to drive close to their desired speeds. Interestingly, the right-hand side of the FD also approximates a straight line, with a slope of about $-14$ m/s.

For the road width of 10.2 m, conventional traffic has three lanes. In lane-based traffic, a typical capacity flow is 2500 veh/km/lane, which amounts to 7500 veh/km for a three-lane road \cite{sasahara2019}. In this work, with an equivalent road width of a three-lane road, more than the double flow has been achieved on the account of lane-free and nudging effects. With lane-free structure, up to five small-size vehicles are seen to occasionally fit in the lateral space, thereby increasing the capacity. Furthermore, nudging also contributes to higher flows, as also observed in \cite{markos2021}. Capacity of of 18332 veh/h (minimum 18213 veh/h and maximum 18446 veh/h) is reached here at a critical density of 200 veh/km. It may also be seen in \Cref{fig_fd} that change of seeds has a moderate to low impact on the flow values at specific densities, with higher dispersion of flows observed near the critical density and near the jam density. Finally, it is interesting to note that no backward-moving jams are observed at any density value.

\subsection{Vehicle-level Results}
Zooming on individual vehicle movements, it should first be mentioned that there are no collisions in all the results reported, although there are very rare occurrences of emergency situations, i.e., cases where the first computed vehicle trajectory includes a collision, hence a new trajectory-planning procedure is triggered with stringer control bounds according to \Cref{sdbc-1,sdbc-2}. Such emergency cases appear only for a very minor percentage of the total number of trajectory planning optimisations. \Cref{tb2} (second row) displays the percentage of occurred emergency cases per traffic density examined. All emergency situations delivered a collision-free vehicle trajectory, after application of the stricter constraints.

%ego-4 no seed
\begin{figure}[!t]
\centering
\includegraphics[scale=0.219]{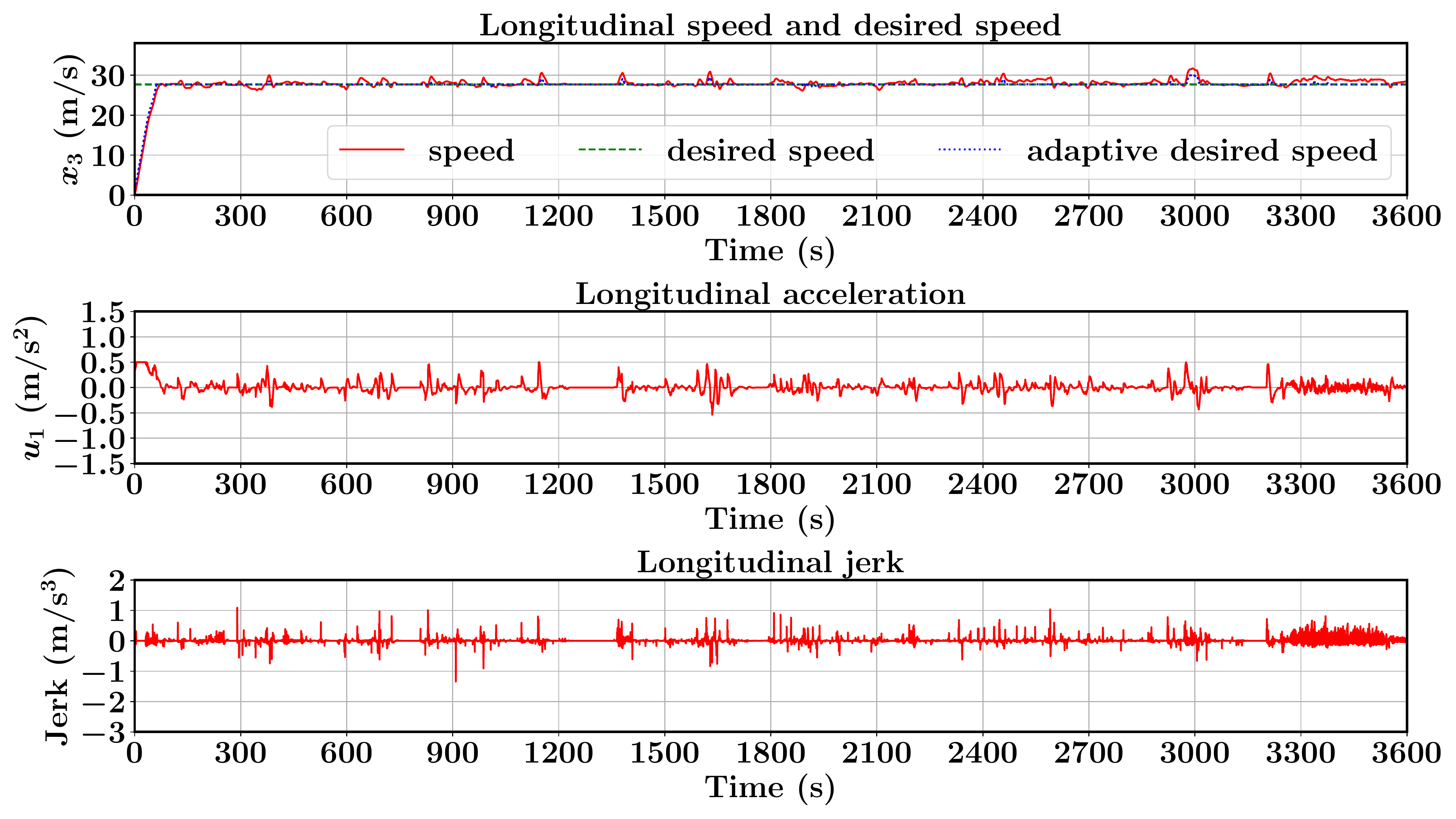}
\caption{Longitudinal movement in 100 veh/km density }\label{long_plot1}
\vspace{-10pt}
\end{figure}

\begin{figure}[!t]
\centering
\includegraphics[scale=0.219]{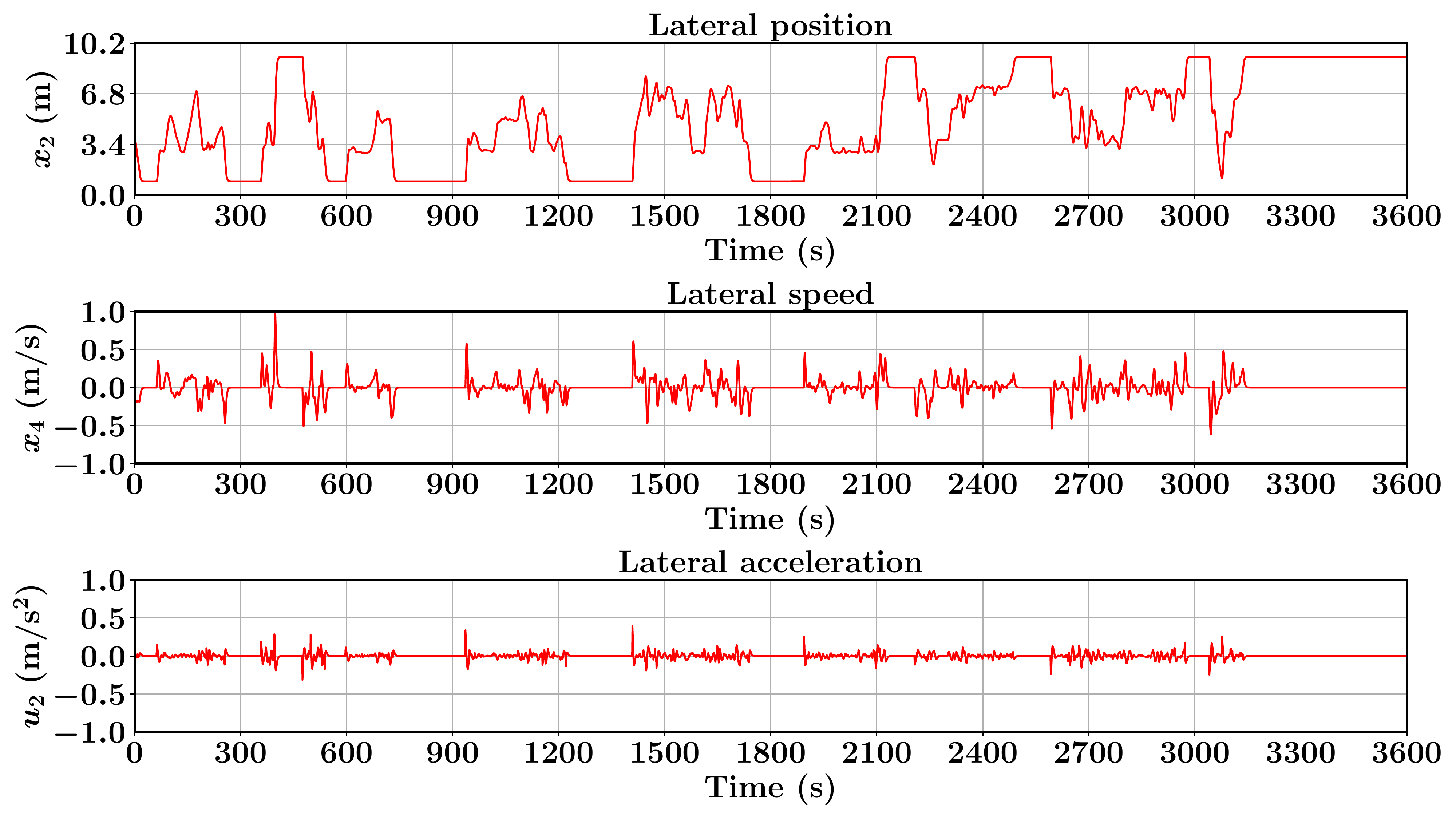}
\caption{Lateral movement in 100 veh/km density}\label{lat_plot1}
\vspace{-10pt}
\end{figure}
\begin{figure}[!t]
\centering
\includegraphics[scale=0.219]{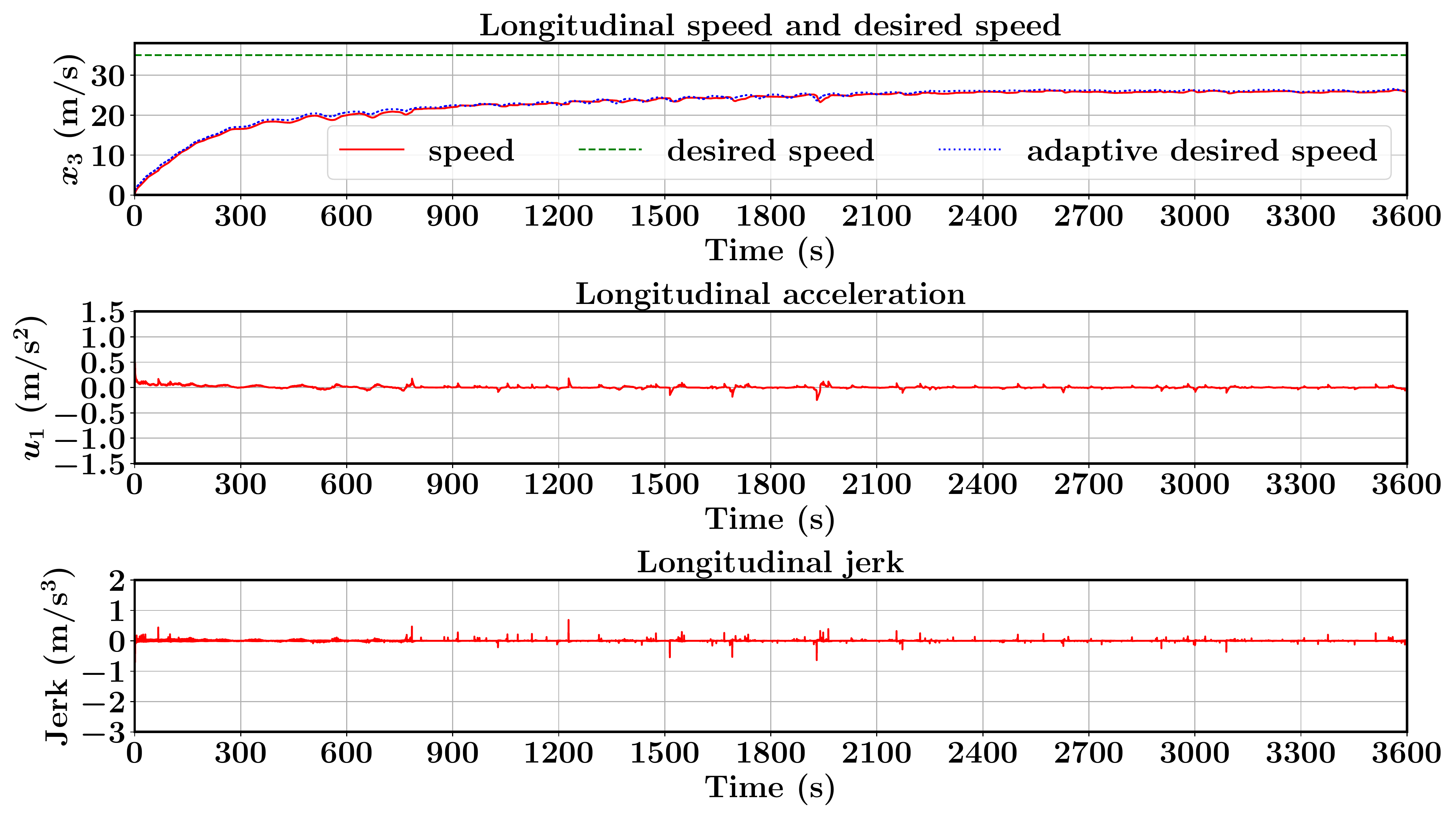}
\caption{Longitudinal movement in 200 veh/km density}\label{long_plot2}
\vspace{-10pt}
\end{figure}
\begin{figure}[!t]
\centering
\includegraphics[scale=0.219]{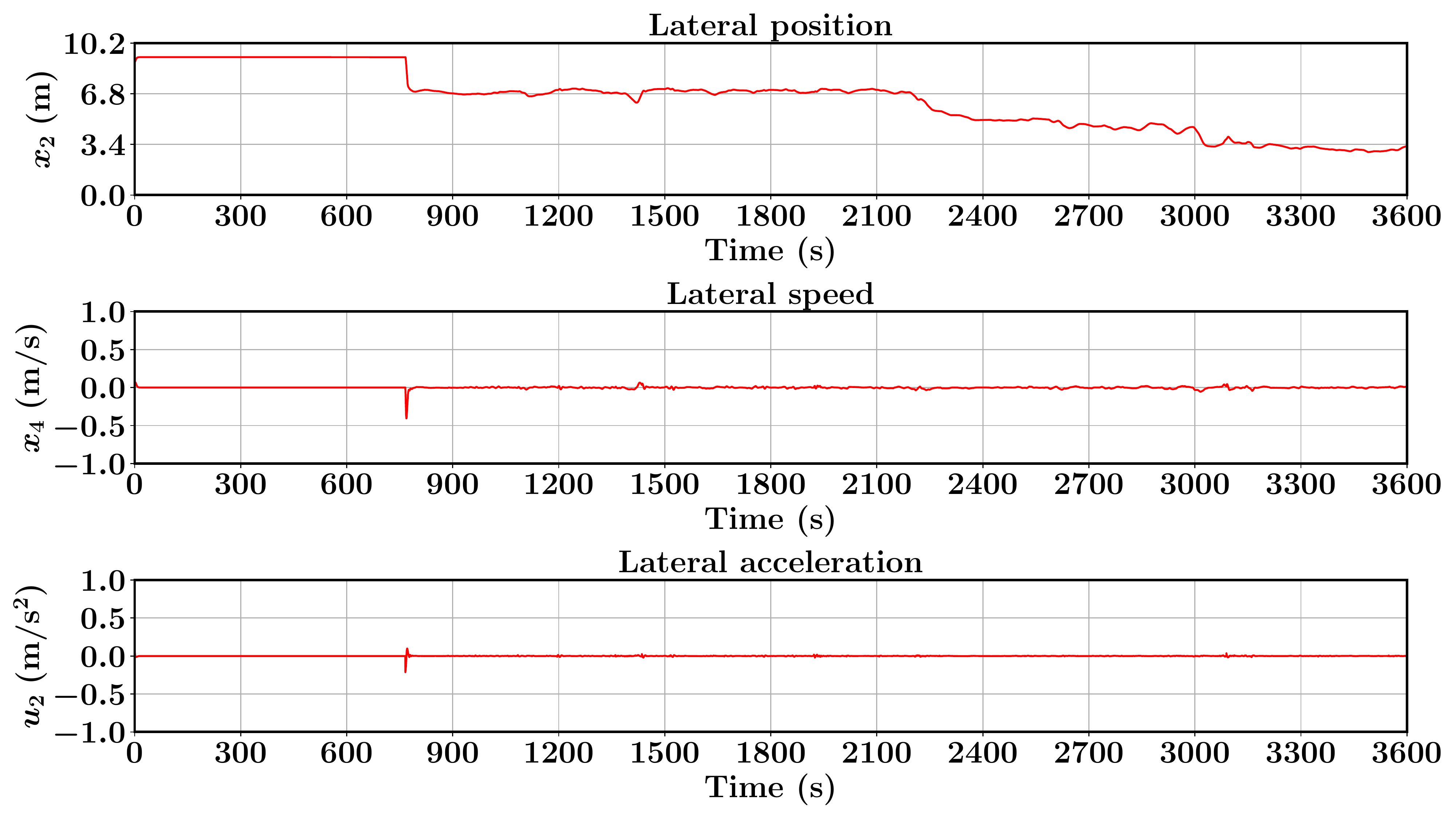}
\caption{Lateral movement in 200 veh/km density}\label{lat_plot2}
\vspace{-10pt}
\end{figure}
\begin{figure}[!t]
\centering
\includegraphics[scale=0.219]{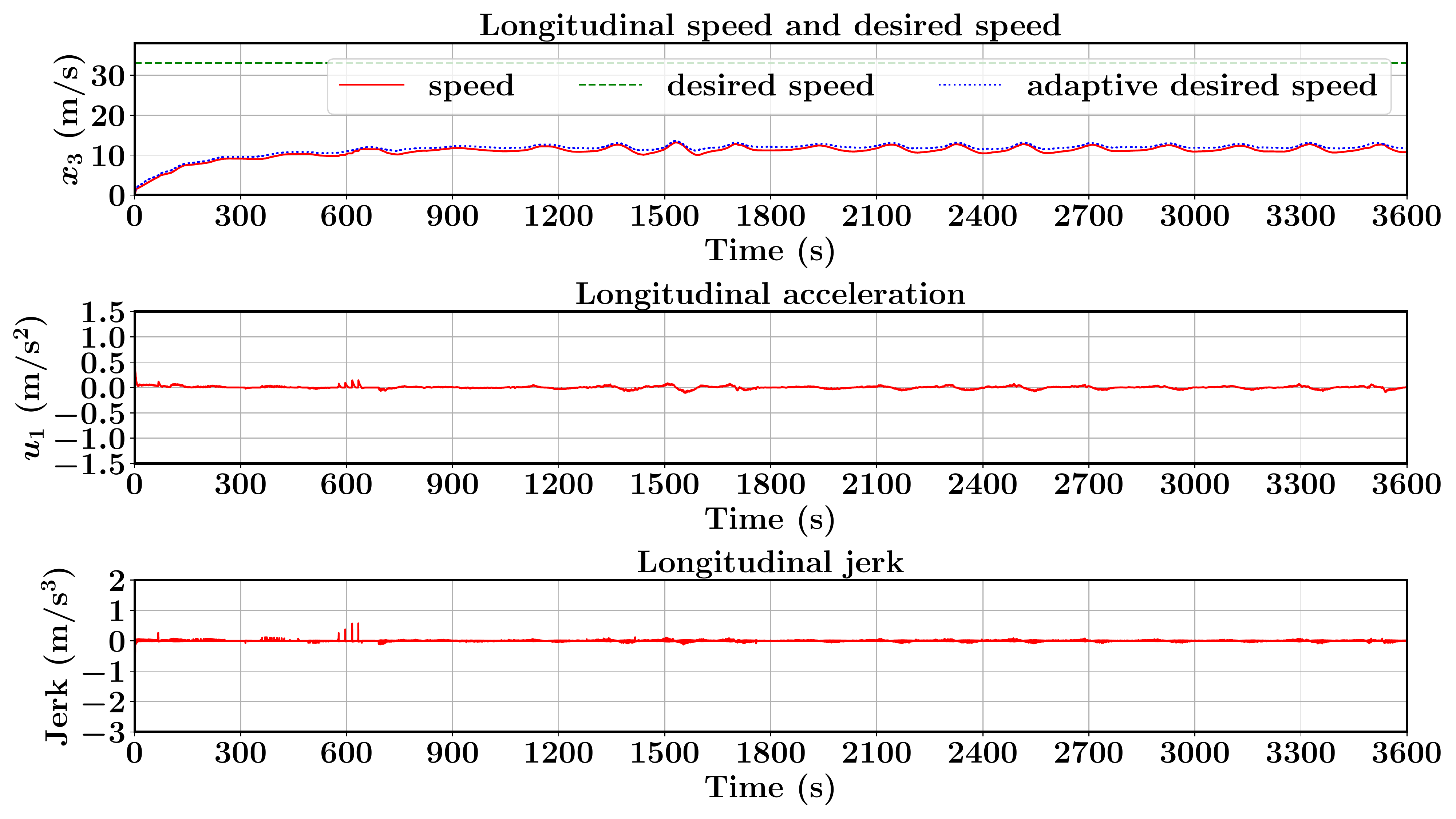}
\caption{Longitudinal movement in 300 veh/km density}\label{long_plot3}
\vspace{-10pt}
\end{figure}
\begin{figure}[!t]
\centering
\includegraphics[scale=0.219]{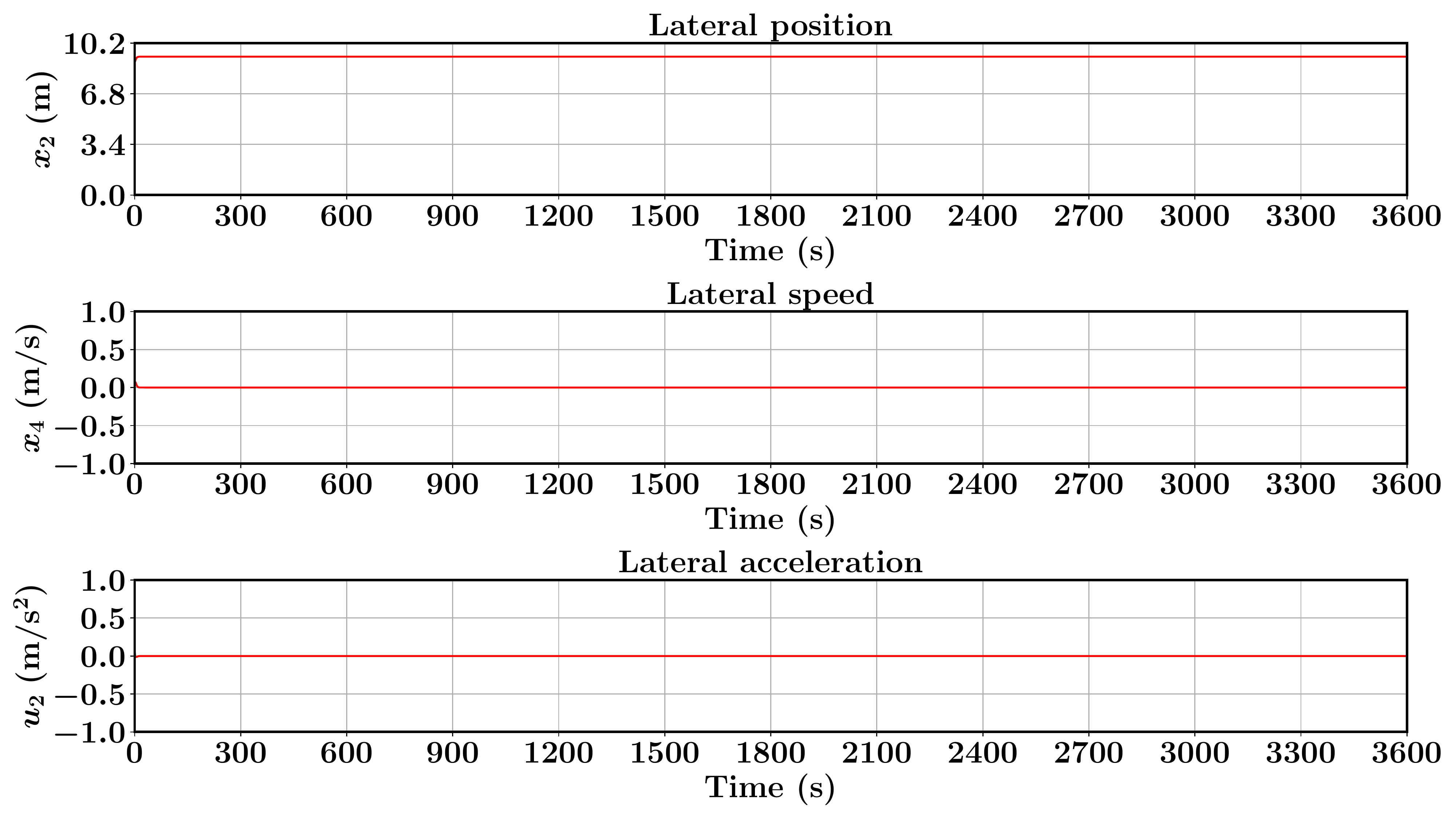}
\caption{Lateral movement in 300 veh/km density}\label{lat_plot3}
\vspace{-10pt}
\end{figure}

To assess the microscopic vehicle behaviour, representative movement trajectories of three vehicles, driving in different densities, are depicted in \cref{long_plot1,lat_plot1,long_plot2,lat_plot2,long_plot3,lat_plot3}. Note that \cref{long_plot1,long_plot2,long_plot3} also display the actual desired longitudinal speed (shown with a green dashed line) and the adaptive desired speed (shown with a blue dotted line). The displayed vehicle trajectories span the whole simulation period. In all the simulations, the vehicles start from zero speed. It can be observed that, at a clearly under-critical density (\Cref{long_plot1}), the vehicle is able to reach its desired longitudinal speed relatively fast, by incrementally approaching the adaptive desired speed at moderate acceleration. Furthermore, the vehicle manages to maintain the desired speed for most of the duration by applying appropriate manoeuvres around encountered slower vehicles. On the other hand, in critical or over-critical densities, as shown in \Cref{long_plot2,long_plot3}, respectively, the vehicle is not able to reach its desired longitudinal speed due to the lack of manoeuvrable space. Instead, the vehicle reaches some quasi-stationary speed, the value of which decreases with increasing densities, as expected, in accordance with the FD of \Cref{fig_fd}. 

\begin{figure}[!t]
\centering
\includegraphics[scale=0.219]{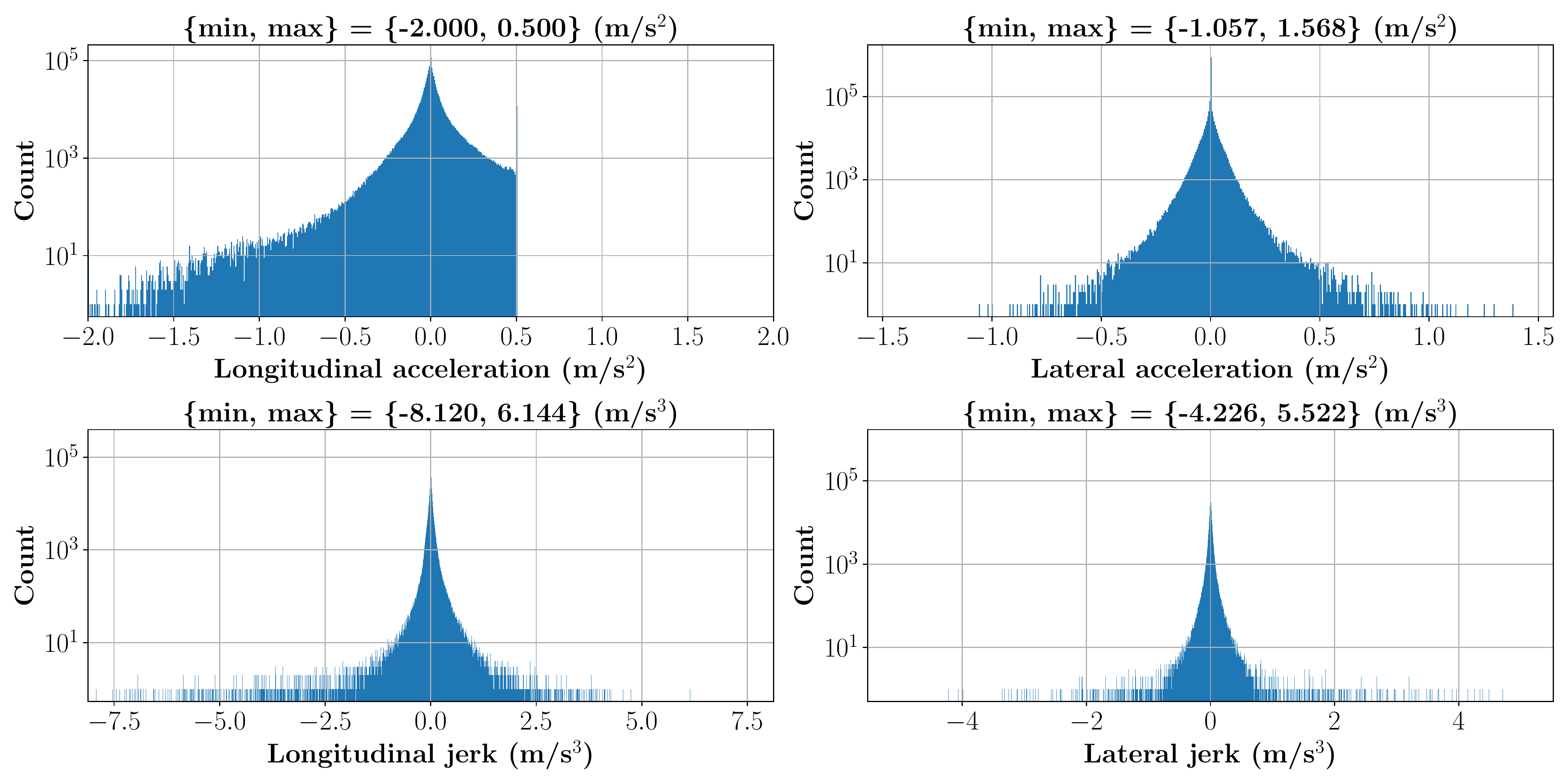}
\caption{Acceleration and jerk histograms for 100 veh/km density}\label{hist_plot1}
\vspace{-10pt}
\end{figure}

\begin{figure}[!t]
\centering
\includegraphics[scale=0.219]{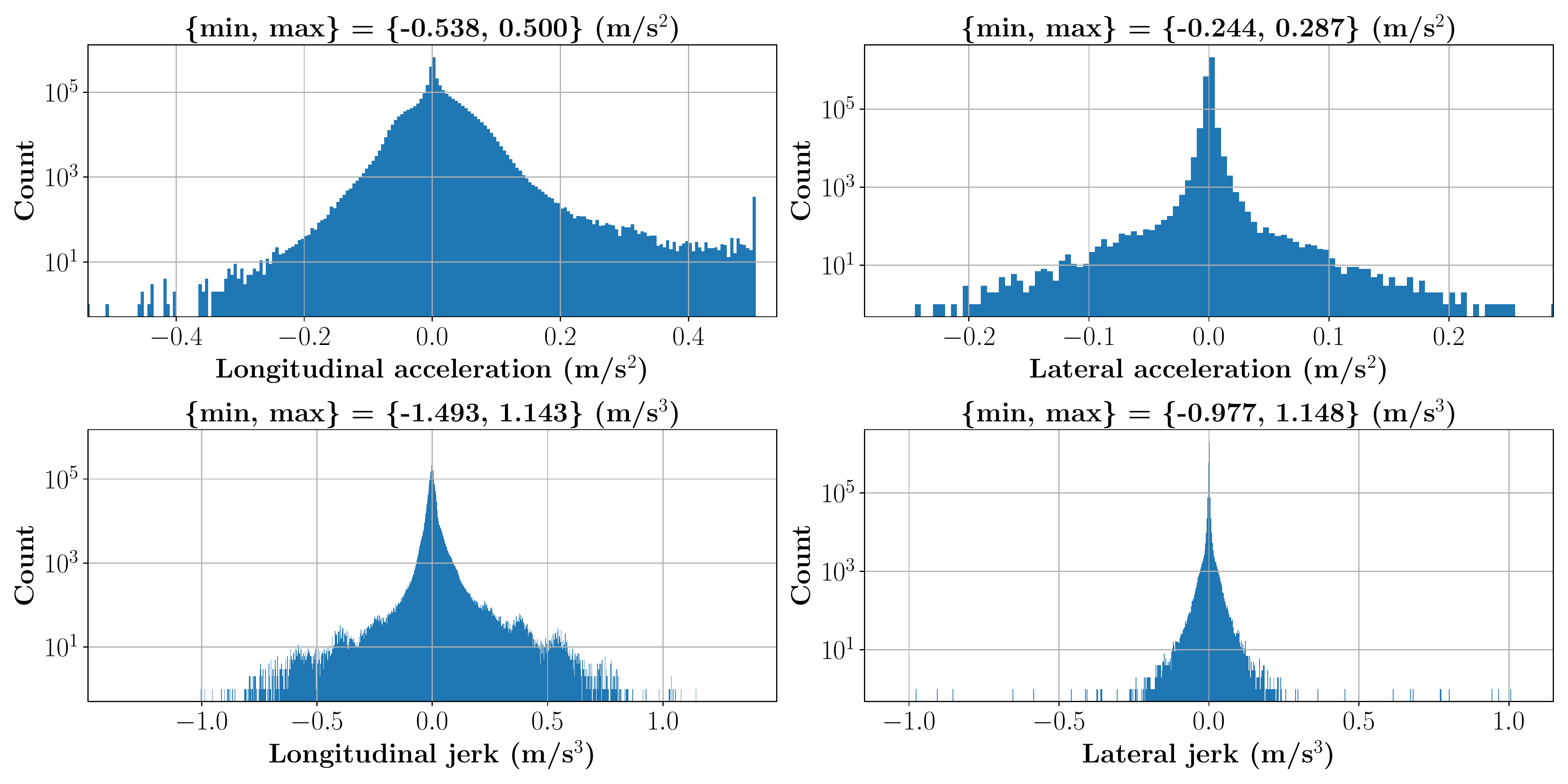}
\caption{Acceleration and jerk histograms for 200 veh/km density}\label{hist_plot2}
\vspace{-10pt}
\end{figure}

\begin{figure}[!t]
\centering
\includegraphics[scale=0.219]{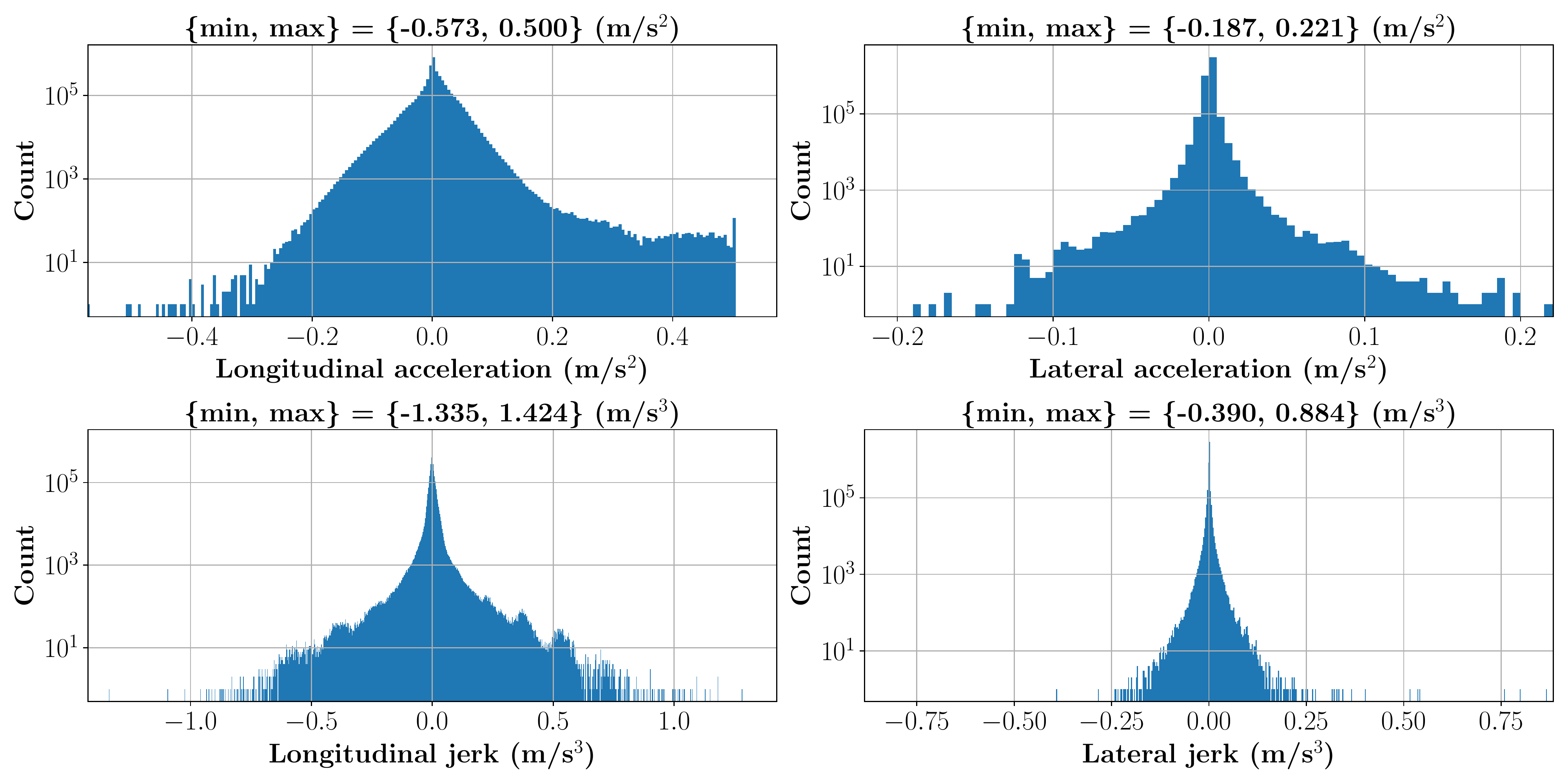}
\caption{Acceleration and jerk histograms for 300 veh/km density}\label{hist_plot3}
\vspace{-10pt}
\end{figure}

It can also be observed that, for all required manoeuvres at all densities, the acceleration and jerk values in the longitudinal direction are well within the comfortable range, which is important for passenger convenience and also better for fuel economy. Histograms of jerk and acceleration values for all vehicles at the same densities, given in \cref{hist_plot1,hist_plot2,hist_plot3}, confirm convenient driving for all vehicles.  

The demonstrated lateral vehicle movement is flexible and differs at different vehicle densities. In under-critical densities, the vehicle is seen in \Cref{lat_plot1} to leverage the complete available lateral space (road width) in its attempt to attain or maintain the desired longitudinal speed. It is interesting to note that lateral manoeuvring is more intensive at periods when the longitudinal speed falls short of the desired speed, which happens due to slower vehicles ahead. In such circumstances, the stronger employed lateral displacements witness the striving of the vehicle to find appropriate opportunities to accelerate longitudinally towards the desired speed. It may also be seen that, at periods, the vehicle is moving exactly on the left or right road boundaries, after having reached those asymptotically, and without ever crossing the boundary, as designed and discussed in \Cref{sdbc-2}. Such efficient vehicle manoeuvring is not only beneficial for vehicle advancement, but also improves the overall traffic efficiency, contributing to the high flows of the FD in \Cref{fig_fd}. While applying these lateral manoeuvres, the vehicle is seen in \Cref{lat_plot1} to maintain moderate or small lateral speed and acceleration.

%ego-4 no seed

In critical or over-critical densities, the lateral movement is limited, as shown in \Cref{lat_plot2,lat_plot3}, due to lack of lateral space, which is occupied by other vehicles. Nevertheless, a few noticeable lateral displacements may be observed, which are usually followed by a slight increase of longitudinal speed. The lateral accelerations and speeds are very mild in all the cases and hence convenient for the passengers, as it can be observed in the corresponding trajectories depicted in \Cref{lat_plot1,lat_plot2,lat_plot3}.

Videos showing the vehicle movement in different densities can be viewed at \href{https://bit.ly/TF-OPP}{\tt \nolinkurl{https://bit.ly/TF-OPP}}. Note that, in the videos, the camera is tracking one vehicle, pointed out by a mouse-pointer, hence all the surrounding vehicles are seen to move with their relative motion with respect to the vehicle being tracked.

%ego-1 no seed

\section{Conclusions}\label{conc}
An optimal trajectory-planning methodology with fixed and state-dependent bounds on control is developed and used to control the vehicle movement on lane-free roads with vehicle nudging. An objective function with multiple sub-objectives for safe, efficient and convenient vehicle movement is designed, along with appropriate state equations and state-dependent control bounds in the frame of an OCP formulation. The sub-objectives are designed to make the EV move smoothly and flexibly in longitudinal and lateral directions without compromising on safety and convenience of the passengers. State-dependent bounds on control are designed, as constraints of the OCP, and are activated, as and when they are needed, in both regular and emergency situations. 

The OCP is appropriately transformed in an NLP problem and solved numerically for a local minimum, using an efficient gradient-based FDA featuring global convergence and polynomial convergence time. The OCP is solved independently for multiple vehicles based on event-triggered MPC for finite time horizons of 8 s in emulated real-time. The vehicle trajectories are updated when there are significant deviations in the predicted trajectories of obstacles or when optimal control for half of the planned trajectory has been already applied. 

Simulations are performed on a ring-road involving hundreds of vehicles with different random initializations. It is observed that the vehicles move efficiently and safely over the full range of possible vehicle densities, while maintaining the passenger convenience. Flow values and critical density values are significantly better than in lane-based traffic.

Ongoing and future work addresses:

\begin{itemize}
\item Introducing on-ramps and off-ramps with appropriate merging and exiting extensions of the present vehicle movement strategy.
\item Testing the performance of the approach at bottlenecks, that may result from high inflow from an on-ramp, both at vehicle level and at traffic level
\item Implementing the internal boundary controller \cite{milad2021} for efficient road utilization.
\item Using nonlinear Ackerman kinematic and the bicycle models \cite{rajamani2012} for roundabouts and intersections.
\end{itemize}

\subsubsection*{Acknowledgement}
This research received funding from the European Research Council under the EU Horizon 2020 Programme / ERC Grant no. 833915, project TrafficFluid, see: \href{https://www.trafficfluid.tuc.gr}{\tt\small \nolinkurl{https://www.trafficfluid.tuc.gr}}

\bibliographystyle{ieeetr}
\bibliography{ref}

%\begin{IEEEbiographynophoto}{All authors}
%\color{red}Biography text and photographs needed. \color{black}
%\end{IEEEbiographynophoto}

\begin{IEEEbiography}[{\includegraphics[width=1in,height=1.25in,clip,keepaspectratio]{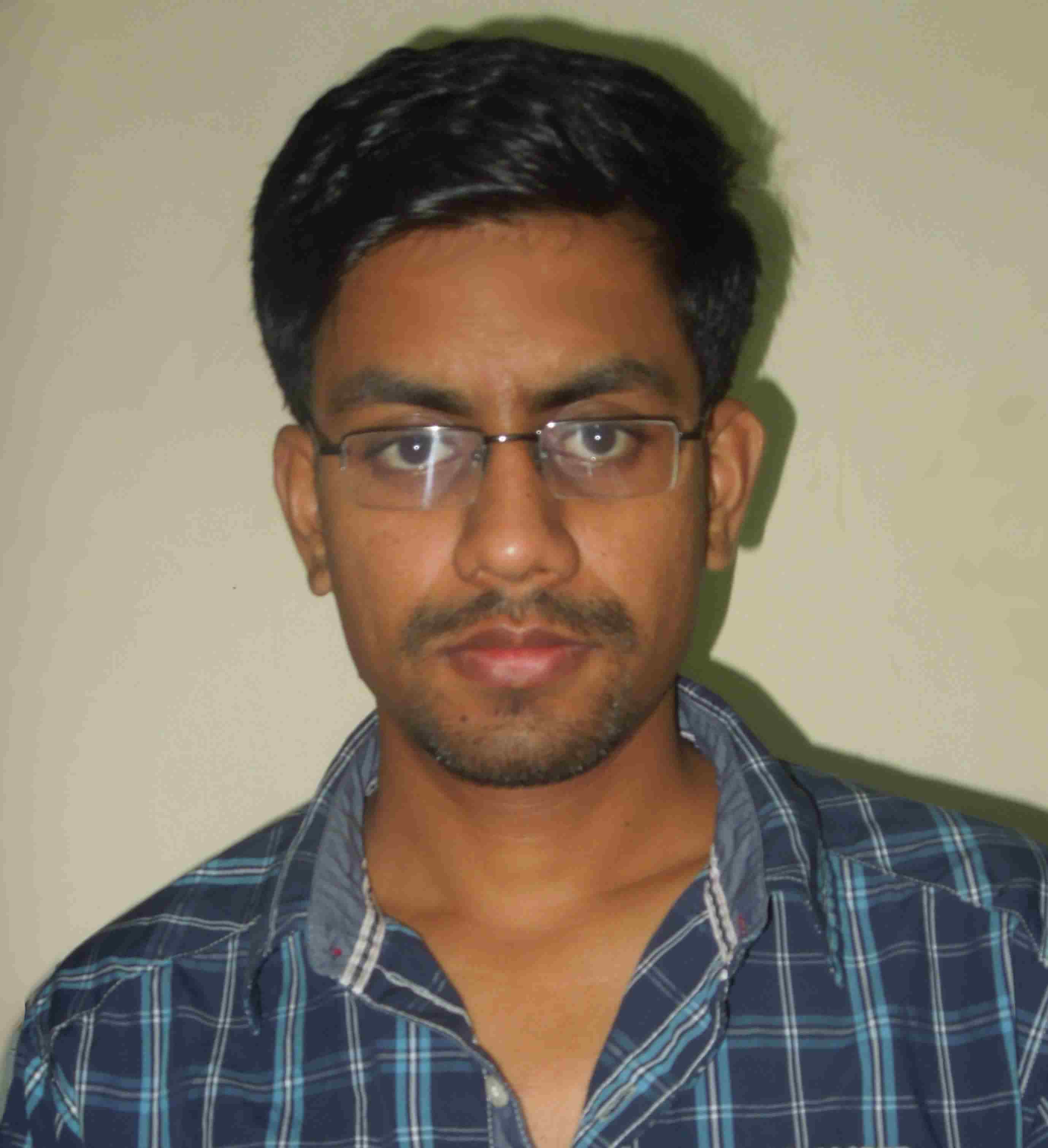}}]{Venkata Karteek Yanumula}
has been a Post Doc Associate with Dynamic Systems \& Simulation Laboratory, Technical University of Crete, Chania, Greece since 2020. He has been a faculty member at Electrical and Instrumentation Engineering Department, Thapar Institute of Engineering and Technology, Patiala, India since 2017. He received PhD in the domain of Systems Control and Automation from Indian Institute of Technology, Guwahati in 2018. His research interests include autonomous vehicles, optimal control, multi-agent systems, and distributed control.\end{IEEEbiography}

\begin{IEEEbiography}[{\includegraphics[width=1in,height=1.25in,clip,keepaspectratio]{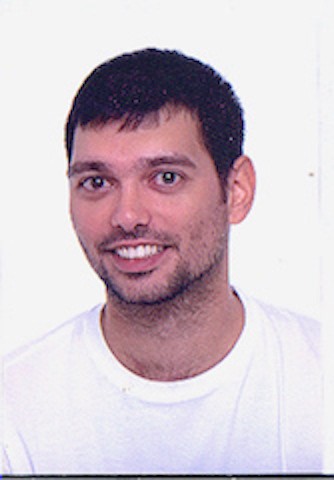}}]{Panagiotis Typaldos}
completed his undergraduate studies in Applied Mathematics at the University of Crete, Heraklion, Greece in 2014 and received the M.Sc. degree in Operational Research from the School of Production Engineering and Management, Technical University of Crete, Chania, Greece, in 2017.

From October 2015, he is a research associate and since September 2017 he has been a PhD student with the Dynamic Systems \& Simulation Laboratory, Technical University of Crete.
\end{IEEEbiography}

\begin{IEEEbiography}[{\includegraphics[width=1in,height=1.25in,clip,keepaspectratio]{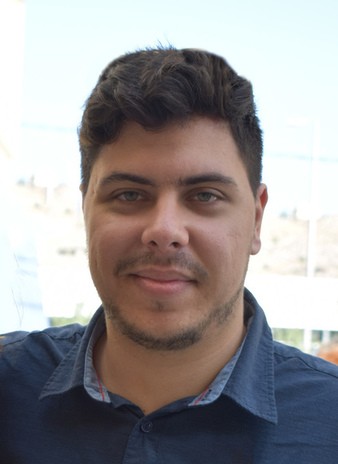}}]{Dimitrios Troullinos}
received the Diploma degree in Electrical and Computer Engineering from the Technical University of Crete, Chania, Greece, in October 2019. His main research interests lie in the area of Multiagent Systems. From November 2019, he is a research associate and since January 2020 he has been a PhD student with the Dynamic Systems \& Simulation Laboratory, Technical University of Crete.\end{IEEEbiography}

\begin{IEEEbiography}[{\includegraphics[width=1in,height=1.25in,clip,keepaspectratio]{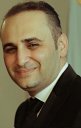}}]{Milad Malekzadeh}
received the B.S. degree in Electrical engineering from the Mazandaran University, Babolsar, Iran, in 2010, and the M.Sc. degree in Control engineering from the Babol Noshirvani University of Technology, Babol, Iran, in 2014. His research interests include observer design, optimal control, and nonlinear dynamics. Since February 2020, he has been a Ph.D. student with the Dynamic Systems \& Simulation Laboratory, Technical University of Crete.\end{IEEEbiography}

\begin{IEEEbiography}[{\includegraphics[width=1in,height=1.25in,clip,keepaspectratio]{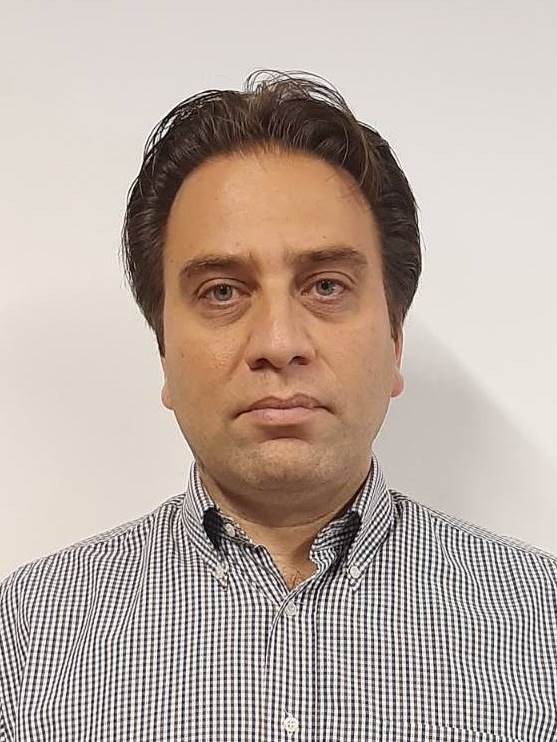}}]{Prof. Ioannis Papamichail}
is the Director of the Dynamic Systems and Simulation Laboratory, at the Technical University of Crete, Chania, Greece. He received the Dipl. Eng. degree in chemical engineering from the National Technical University of Athens, in 1998 and the M.Sc. degree in process systems engineering and the Ph.D. degree in chemical engineering from Imperial College London, in 1999 and 2002, respectively.

From 1999 to 2002, he was a Research Assistant with the Center for Process Systems Engineering, Imperial College London. He joined the Technical University of Crete in 2004 and has served, since then, at all academic ranks. In 2010, he was a Visiting Scholar with the University of California, Berkeley, CA, USA. He is the author of several technical papers in scientific journals and conference proceedings. His main research interests include automatic control and optimization theory and applications to traffic and transportation systems.

Dr. Papamichail is an Associate Editor for IEEE Transactions on Intelligent Transportation Systems and a Member of the Editorial Advisory Board for Transportation Research Part C. He received the 1998 Eugenidi Foundation Scholarship for Postgraduate Studies and the 2010 Transition to Practice Award from the IEEE Control Systems Society.
\end{IEEEbiography}

\begin{IEEEbiography}[{\includegraphics[width=1in,height=1.25in,clip,keepaspectratio]{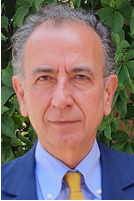}}]{Markos Papageorgiou}
(Life Fellow, IEEE) was a Professor of Automation with the Technical University of Munich, Germany, from 1988 to 1994. Since 1994 he has been a Professor (since 2021 Professor Emeritus) at the Technical University of Crete, Chania, Greece. Since 2021 he has been a Professor at Ningbo University, China. He was a Visiting Professor with the Politecnico di Milano, the \'{E}cole Nationale des Ponts et Chauss\'{e}es, MIT, the Sapienza University of Rome, and Tsinghua University, and a Visiting Scholar with UC Berkeley. His research interests include automatic control and optimisation theory and applications to traffic and transportation systems, water systems, and further areas. He is a fellow of IFAC. He received several distinctions and awards, including the 2020 IEEE Transportation Technologies Award and two ERC Advanced Investigator Grants.\end{IEEEbiography}

\end{document}